\documentclass[12pt]{article}
\usepackage[margin=1.9cm]{geometry}
\usepackage{graphicx}
\usepackage{cite}
\usepackage{float}
\usepackage{multirow}
\usepackage{titlesec}

\newcommand{\mysection}{\setcounter{equation}{0}\section}
\renewcommand{\theequation}{\thesection.\arabic{equation}}
\def\beq{\begin{equation}}
\def\eeq{\end{equation}}
\def\beqa{\begin{eqnarray}}
\def\eeqa{\end{eqnarray}}

\begin{document}

\begin{center}
{\Large \bf N$^3$LO soft-gluon corrections for $Z\gamma$ production}
\end{center}

\vspace{2mm}

\begin{center}
{\large Nikolaos Kidonakis and Kaan \c{S}im\c{s}ek} \\

\vspace{2mm}

{\sl Department of Physics, Kennesaw State University, Kennesaw, Georgia 30144, USA}

\end{center}

\begin{abstract}
We calculate soft-gluon corrections for $Z\gamma$ production through next-to-next-to-next-to-leading-order (N$^3$LO) in one-particle-inclusive (1PI) kinematics. We use these results to produce approximate N$^3$LO (aN$^3$LO) QCD  predictions for $Z\gamma$ production cross sections in proton-proton collisions at LHC energies. Differential cross sections in photon transverse momentum $p_T^\gamma$ and rapidity $y^\gamma$ are computed at 13 TeV and 13.6 TeV collider energies. Higher-order $K$-factors are examined systematically, and scale and PDF uncertainties are quantified at each perturbative order. Soft-gluon corrections are found to be significant in the high-$p_T$ region, with good perturbative convergence observed at aN$^3$LO.
\end{abstract}

\mysection{Introduction}

The associated production of a $Z$ boson and a photon at the Large Hadron Collider (LHC) is an important probe of the electroweak sector of the Standard Model (SM). The process $pp \to Z\gamma$ proceeds through quark-antiquark annihilation at leading order and provides direct sensitivity to the gauge structure of the electroweak interaction, including the $ZZ\gamma$ and $Z\gamma\gamma$ neutral gauge-boson vertices, which are absent at tree level in the SM. Precise theoretical predictions for $Z\gamma$ production are therefore essential for testing the SM at high energies, constraining anomalous neutral triple gauge couplings, and providing a well-understood baseline for searches for physics beyond the Standard Model.

Because the $ZZ\gamma$ and $Z\gamma\gamma$ vertices vanish at tree level in the SM, any observed deviation in $Z\gamma$ production is a direct sign of new physics. The ATLAS collaboration has searched for anomalous neutral triple gauge couplings in $Z\gamma$ production at 13 TeV collider energy using the neutrino decay channel of the $Z$ boson~\cite{ATLAS2018}. The CMS collaboration has performed a similar search in the same channel~\cite{CMS2026}. In the SM effective field theory, neutral triple gauge couplings first appear from dimension-eight operators~\cite{Degrande2013}. Precise SM predictions for the $Z\gamma$ cross section and its kinematic distributions are needed to interpret these searches and to place robust limits on new physics.

Fixed-order perturbative QCD calculations for $Z\gamma$ production have reached a high level of maturity. Next-to-leading-order (NLO) corrections were computed in Ref.~\cite{Ohnemus1993}, and next-to-next-to-leading order (NNLO) predictions have been obtained using the $q_T$-subtraction formalism~\cite{GKRT2013,GKR2015}. These results have been compared against measurements by the ATLAS~\cite{ATLAS2019} and CMS~\cite{CMS2026} collaborations at 13 TeV energy, demonstrating good agreement within theoretical and experimental uncertainties. Despite this progress, fixed-order predictions suffer from residual scale uncertainties that grow with the photon transverse momentum $p_T^\gamma$, driven by the increasing dominance of threshold logarithms in the partonic cross section.

Near partonic threshold, where the invariant mass of the final state approaches the partonic center-of-mass energy, there is limited phase space for additional gluon radiation. Thus, soft-gluon logarithmic contributions of the form $\ln^k(s_4/m_Z^2)/s_4$, with $s_4$ a partonic-threshold variable, appear at each perturbative order, and they can be formally resummed under integral transforms. The expansion of the resummed cross section provides analytical expressions for higher-order soft-gluon contributions which are numerically overwhelmingly dominant in $Z \gamma$ production and, thus, systematically improve upon lower-order results.

In this paper, we employ the soft-gluon formalism of Refs. \cite{Sterman1987,NKGS1996,NKGS1997,KOS1998,LOS1998,Kidonakis2007,Kidonakis2010,NKRG2014,Kidonakis2017,MFNK2020,NKNY2022,NKAT2024,NKCF2024,NKAT2025} to $Z\gamma$ production using one-particle-inclusive (1PI) kinematics, focusing on the photon as the observed particle. We calculate the soft-gluon corrections through next-to-next-to-next-to-leading order (N$^3$LO). By adding these higher-order corrections to the NLO result, we construct approximate NNLO (aNNLO) and approximate N$^3$LO (aN$^3$LO) differential cross sections, and we present numerical predictions for the photon transverse momentum $p_T^\gamma$ and rapidity $y^\gamma$ distributions at LHC energies of 13 TeV and 13.6 TeV. Fixed-order benchmarks at LO and NLO are obtained using \textsc{MadGraph5\_aMC@NLO}~\cite{Alwalletal2014, Frederixetal2018}, and $K$-factors are examined systematically to assess the convergence of the perturbative series.

This paper is organized as follows. In Section 2, we present the soft-gluon resummation formalism for $Z\gamma$ production. In Section 3, we describe our numerical setup and present differential cross sections, $K$-factors, and scale and PDF uncertainties for the photon transverse momentum and rapidity distributions at LO, NLO, aNNLO, and aN$^3$LO. We give our conclusions in Section 4. Explicit analytic expressions for the soft-gluon corrections through N$^3$LO are presented in the Appendix.

\mysection{Soft-gluon resummation for $Z \gamma$ production}

In this section, we present the resummation formalism for the calculation of soft-gluon corrections in $\gamma Z$ production through N$^3$LO. We work in 1PI kinematics with the photon as the observed particle. We use the general framework as developed and described in Refs. \cite{Sterman1987,NKGS1996,NKGS1997,KOS1998,LOS1998,Kidonakis2007,Kidonakis2010,NKRG2014,Kidonakis2017,MFNK2020,NKNY2022,NKAT2024,NKCF2024,NKAT2025}. The soft-gluon corrections arise from partial cancellations of infrared divergences between virtual contributions and real-emission contributions from low-energy (soft) gluons close to partonic threshold for the production of a $\gamma Z$ final state.

The leading-order partonic processes are  
\beq
q (p_a)\, + {\bar q} (p_b) \to \gamma (p_1) \, + \, Z (p_2)  \, ,
\eeq
where $q$ and ${\bar q}$ are quarks and antiquarks in the protons. We define $s=(p_a+p_b)^2$, $t=(p_a-p_1)^2$, $t_1=t-m_Z^2$, $u=(p_b-p_1)^2$, $u_1=u-m_Z^2$, as well as a partonic threshold variable $s_4=s+t+u-m_Z^2$, with $m_Z$ the $Z$-boson mass. Near partonic threshold, $s_4 \to 0$, and the soft-gluon corrections appear in the perturbative series as logarithms of $s_4$, i.e. $[(\ln^k(s_4/m_Z^2))/s_4]_+$, with $0 \le k \le 2n-1$ at $n$th order in the strong coupling $\alpha_s$.
 
The resummation of soft-gluon corrections follows from the factorization properties of the differential cross section under Laplace transforms into a hard function $H_{q{\bar q} \to \gamma Z}$, a soft function $S_{q{\bar q} \to \gamma Z}$, and parton distribution functions (PDF), and the renormalization group evolution of these functions. We outline the steps involved in soft-gluon resummation below.

First, we write the differential cross section in proton-proton collisions, $d\sigma_{pp \to \gamma Z}$, as a convolution in the form 
\beq
d\sigma_{pp \to \gamma Z }=\sum_{q,{\bar q}} \; 
\int dx_a \, dx_b \,  \phi_{q/p}(x_a, \mu_F) \, \phi_{{\bar q}/p}(x_b, \mu_F)  \, 
d{\hat \sigma}_{q{\bar q} \to \gamma Z}(s_4, \mu_F, \mu_R)  \, ,
\label{factorized}
\eeq
where $\mu_F$ is the factorization scale, $\mu_R$ is the renormalization scale, $d{\hat \sigma}_{q{\bar q} \to \gamma Z}$ is the differential partonic cross section, and $\phi_{q/p}$, $\phi_{{\bar q}/p}$ are PDF with $x_a$, $x_b$ the momentum fractions of partons $q$, ${\bar q}$, respectively. 

The LO differential partonic cross section is
\beq
\frac{d{\hat{\sigma}}^{(0)}_{q{\bar q} \to \gamma Z}}{dt_1 \, du_1}  = F^{(0)}_{q{\bar q} \to \gamma Z} \, \delta(s_4)
\label{LO}
\eeq
where
\beq
F^{(0)}_{q{\bar q} \to \gamma Z}=\frac{e^4 Q^2}{32 \pi N_c \, s^2} \frac{\left(c_V^2+c_A^2\right)}{\sin^2\theta_w \cos^2 \theta_w} \left(\frac{t}{u}+\frac{u}{t}+\frac{2 m_Z^2 s}{t u} \right) \, .
\eeq 
Here $Q=2/3$, $c_V=1/2 -4/3 \, \sin^2\theta_w$, and $c_A=1/2$ for $q=u$, $c$, $t$, while $Q=-1/3$, $c_V=-1/2 +2/3 \, \sin^2\theta_w$, and $c_A=-1/2$  for $q=d$, $s$, $b$.

We also define the hadronic variables $S=(P_a+P_b)^2$, $T=(P_a-p_1)^2$, and $U=(P_b-p_1)^2$, where $P_a$ and $P_b$ are the momenta of the two colliding protons, as well as $S_4=S+T+U-m_Z^2$. Since $p_a=x_a P_a$ and $p_b=x_b P_b$, we have the relations $s=x_a x_b S$, $t=x_a T$, $u=x_b U$, and
\beq
\frac{S_4}{S}=\frac{s_4}{s}-(1-x_a) \frac{u_1}{s}-(1-x_b) \frac{t_1}{s} -\frac{(1-x_a)(1-x_b)}{s} m_Z^2\, .
\label{S4}
\eeq
The last term in Eq. (\ref{S4}) is ${\cal O}((1-x_a)(1-x_b))$ and can, thus,  be ignored in the large-$x$ threshold limit $x_a \to 1$ and $x_b \to 1$.

We then consider the parton-parton cross section $d\sigma_{q{\bar q} \to \gamma Z}$, which is of the same form as Eq.~(\ref{factorized}) but with incoming partons instead of hadrons \cite{NKGS1996,NKGS1997,LOS1998},  
\beq
d\sigma_{q{\bar q} \to \gamma Z}(S_4)=
\int dx_a \, dx_b \,  \phi_{q/q}(x_a) \, \phi_{{\bar q}/{\bar q}}(x_b) \, 
d{\hat \sigma}_{q{\bar q} \to \gamma Z}(s_4) \, ,
\label{factphi}
\eeq
and we define its Laplace transform as 
\beq
d{\tilde \sigma}_{q{\bar q} \to \gamma Z}(N)=\int_0^{S_{4 \, \rm max}} 
\frac{dS_4}{S} \,  e^{-N S_4/S} \, d\sigma_{q{\bar q} \to \gamma Z}(S_4) \, . 
\label{cstr}
\eeq 

Using the relation for $S_4/S$ in Eq. (\ref{S4}), we rewrite Eq. (\ref{cstr}) as  
\beqa
d{\tilde \sigma}_{q{\bar q} \to \gamma Z}(N) &=& \int_0^1 dx_a e^{-N_a (1-x_a)} 
\phi_{q/q}(x_a) \int_0^1 dx_b e^{-N_b (1-x_b)} \phi_{{\bar q}/{\bar q}}(x_b)
\int_0^{s_{4 \, {\rm max}}} \frac{ds_4}{s} e^{-N s_4/s} d{\hat \sigma}_{q{\bar q} \to \gamma Z}(s_4)
\nonumber \\ 
&=& {\tilde \phi}_{q/q}(N_a) \, {\tilde \phi}_{{\bar q}/{\bar q}}(N_b) \, 
d{\tilde{\hat \sigma}}_{q{\bar q} \to \gamma Z}(N) \, ,
\label{fac}
\eeqa
where $N_a=N(-u_1/s)$ and $N_b=N(-t_1/s)$.

Next, we introduce a refactorization of the cross section via the functions $H_{q{\bar q} \to \gamma Z}$, $S_{q{\bar q} \to \gamma Z}$, $\psi_{q/q}$, and $\psi_{{\bar q}/{\bar q}}$ \cite{NKGS1996,NKGS1997,LOS1998}. The short-distance hard function $H_{q{\bar q} \to \gamma Z}$ is infrared safe while the soft function $S_{q{\bar q} \to \gamma Z}$ describes wide-angle soft gluon emission. The coupling of the soft gluons to the partons is described via the introduction of Wilson lines, which are ordered exponentials of the gluon gauge field. The functions $\psi_{q/q}$ and $\psi_{{\bar q}/{\bar q}}$ describe collinear emission from the incoming partons \cite{Sterman1987,NKGS1996,NKGS1997,LOS1998}. The refactorized form of the cross section is  
\beqa
d{\sigma}_{q{\bar q} \to \gamma Z}&=&\int dw_a \, dw_b \, dw_S \, \psi_{q/q}(w_a) \, \psi_{{\bar q}/{\bar q}}(w_b) 
\nonumber \\ && \times
H_{q{\bar q} \to \gamma Z} \, \, 
S_{q{\bar q} \to \gamma Z}\left(\frac{w_S \sqrt{s}}{\mu_F} \right) \; 
\delta\left(\frac{S_4}{S}-w_S+w_a\frac{u_1}{s}
+w_b \frac{t_1}{s}\right)
\label{refact}
\eeqa
where the $w$'s are dimensionless weights, with $w_a$ and $w_b$ for $\psi_{q/q}$ and $\psi_{{\bar q}/{\bar q}}$, respectively, and $w_S$ for $S_{q{\bar q} \to \gamma Z}$. 

By taking a Laplace transform of Eq. (\ref{refact}), we find
\beqa
d{\tilde \sigma}_{q{\bar q} \to \gamma Z}(N)&=& 
\int_0^1 dw_a \, e^{-N_a w_a} \, \psi_{q/q}(w_a) \int_0^1 dw_b \, e^{-N_b w_b} \, \psi_{{\bar q}/{\bar q}}(w_b)
\nonumber \\ && \times \, \,
H_{q{\bar q} \to \gamma Z}  \int_0^1 dw_s \, e^{-N w_s} \, S_{q{\bar q} \to \gamma Z}\left(\frac{w_s\sqrt{s}}{\mu_F} \right)
\nonumber \\ &=& 
{\tilde \psi}_{q/q}(N_a) \, \, {\tilde \psi}_{{\bar q}/{\bar q}}(N_b) \, \, 
H_{q{\bar q} \to \gamma Z} \, \, 
{\tilde S}_{q{\bar q} \to \gamma Z}\left(\frac{\sqrt{s}}{N \mu_F} \right) \, .
\label{refac}
\eeqa

Then, comparing Eqs. (\ref{fac}) and (\ref{refac}), we find an expression for the partonic cross section in Laplace transform space,
\beq
d{\tilde{\hat \sigma}}_{q{\bar q} \to \gamma Z}(N)=
\frac{{\tilde \psi}_{q/q}(N_a) \, {\tilde \psi}_{{\bar q}/{\bar q}}(N_b)}
{{\tilde \phi}_{q/q}(N_a) \, {\tilde \phi}_{{\bar q}/{\bar q}}(N_b)} \, \,  
H_{q{\bar q} \to \gamma Z} \, \, 
{\tilde S}_{q{\bar q} \to \gamma Z}\left(\frac{\sqrt{s}}{N \mu_F} \right) \, .
\label{sigN}
\eeq

The renormalization-group evolution of ${\tilde \psi}_{q/q}/{\tilde \phi}_{q/q}$, $ {\tilde \psi}_{{\bar q}/{\bar q}}/{\tilde \phi}_{{\bar q}/{\bar q}}$, and ${\tilde S}_{q{\bar q} \to \gamma Z}$ results in exponentiation, i.e. resummation, of the soft-gluon contributions. The resummed cross section is given in terms of the Laplace transform variable $N$ by 
\beqa
d\hat{\sigma}^{\rm resum}_{q{\bar q} \to \gamma Z}(N)&=&   
\exp\left[E_q (N_a)+ E_{\bar q} (N_b)\right] \; 
\exp \left[2\int_{\mu_F}^{\sqrt{s}} \frac{d\mu}{\mu}\; 
\left(\gamma_{q/q}\left({\tilde N}_a,\alpha_s(\mu)\right)
+\gamma_{{\bar q}/{\bar q}}\left({\tilde N}_b,\alpha_s(\mu)\right)\right)
\right]
\nonumber \\ && \quad
\times \,\, H_{q{\bar q} \to \gamma Z}\left(\alpha_s({\sqrt s})\right) \;
{\tilde S}_{q{\bar q} \to \gamma Z}\left(\alpha_s({\sqrt s}/N)\right) \, .
\label{resdsigma}
\eeqa
The first exponent of Eq. (\ref{resdsigma}) is,
\beq
E_q(N_a)=
\int^1_0 dz \frac{z^{N_a-1}-1}{1-z}\;
\left \{\int_1^{(1-z)^2} \frac{d\lambda}{\lambda}
A_q \left(\alpha_s(\lambda s)\right)
+D_q \left[\alpha_s((1-z)^2 s)\right]\right\} 
\label{Eexp}
\eeq
where $A_q$ is the lightlike cusp anomalous dimension, and $D_q$ is a soft-gluon exponent. An analogous expression holds for $E_{\bar q}(N_b)$. In the second exponent of Eq. (\ref{resdsigma}), the quantity $\gamma_{q/q}$ is the moment-space anomalous dimension of the ${\overline {\rm MS}}$ density $\phi_{q/q}$, and similarly for  $\gamma_{{\bar q}/{\bar q}}$. We use the same notation and expressions for the exponents as in Ref. \cite{NKAT2025}.

We expand Eq. (\ref{resdsigma}) at fixed order to N$^3$LO, and then we invert back to momentum space to derive results for the soft-gluon corrections to the physical cross section. Explicit analytical expressions are provided in the Appendix.

\mysection{Numerical analysis}

LO and NLO predictions are obtained using {\small \sc MadGraph5\_aMC@NLO}~\cite{Alwalletal2014,Frederixetal2018}. The aNNLO and aN$^3$LO predictions are computed using our in-house threshold resummation code based on the soft-gluon framework of Refs. \cite{Sterman1987,NKGS1996,NKGS1997,KOS1998,LOS1998,Kidonakis2007,Kidonakis2010,NKRG2014,Kidonakis2017,MFNK2020,NKNY2022,NKAT2024,NKCF2024,NKAT2025} using 1PI kinematics.

Electroweak couplings are derived from the $\{G_F, m_Z, m_W\}$ input scheme with
\beqa
    G_F &=& 1.1663788 \times 10^{-5} \; {\rm GeV}^{-2}, \\
    m_Z &=& 91.1880 \; {\rm GeV}, \\
    m_W &=& 80.3692 \; {\rm GeV},
\eeqa
with $\Gamma_Z = 2.4955$ GeV. The weak mixing angle is determined by $c_W = m_W/m_Z$ and $s_W = \sqrt{1 - c_W^2}$, and the electromagnetic coupling $\alpha$ is derived from $G_F$, $m_W$, and $s_W$ via $\alpha = \sqrt{2}\,G_F m_W^2 s_W^2/\pi$. The strong coupling $\alpha_s$ is taken from the PDF set used at each order, evaluated at $\mu = m_Z$. At NLO QCD, photons are isolated using the smooth cone criterion of Ref.~\cite{Frixione1998} with cone radius $R_0 = 0.4$, exponent $n = 1$, and energy fraction $\epsilon_\gamma = 0.5$.

A minimum photon transverse momentum $p_T^\gamma > 20$~GeV is imposed in all calculations. No additional requirements are placed on lepton kinematics, photon rapidity, or geometric separations, as these are not accessible within the 1PI formulation. The collider energy is taken to be 13 TeV and 13.6 TeV, corresponding to LHC Run~2 and Run~3 conditions, respectively.

The central renormalization and factorization scales are set to $\mu_R = \mu_F = m_Z$, with uncertainties estimated via 3-point variation over $\mu_{R,F} \in \{m_Z/2,\, m_Z,\, 2m_Z\}$ with $\mu_R = \mu_F$. All predictions are computed using both the MSHT20 nnlo ~\cite{MSHT20nnlo} and MSHT20 an3lo ~\cite{MSHT20an3lo} PDF sets at every perturbative order so that differences between orders reflect genuine higher-order QCD corrections rather than PDF evolution effects. PDF uncertainties are evaluated using the full MSHT20 Hessian error sets via supplementary {\small \sc MadGraph5\_aMC@NLO} runs at LO and NLO, and the resulting fractional uncertainties are applied to the central predictions at each order.

For presentation of differential distributions, we adopt the $p_T^\gamma$ and $|y^\gamma|$ binning of the ATLAS $Z\gamma$ measurement at $\sqrt{s} = 13$~TeV with 139~fb$^{-1}$~\cite{ATLAS2019}. For a massless photon, $E_T^\gamma = p_T^\gamma$ and $\eta^\gamma = y^\gamma$, so the ATLAS distributions correspond directly to our observables. The $p_T^\gamma$ distribution is presented in 12 bins:
\beq
    p_T^\gamma \in [30, 40, 50, 60, 70, 80, 100, 120, 150, 200, 300, 500, 1200] \, {\rm GeV},
\eeq
and the $|y^\gamma|$ distribution in 8 bins:
\beq
    |y^\gamma| \in [0, 0.3, 0.6, 0.9, 1.2, 1.5, 1.8, 2.1, 2.4].
\eeq
We note that this choice is made purely for presentation and does not imply a direct comparison to the experimental fiducial cross section.

With this setup fixed, we compute the total fiducial cross section and the differential distributions $d\sigma/d p_T^\gamma$ and $d\sigma/d|y^\gamma|$ at LO, NLO, aNNLO, and aN$^3$LO. We note that the soft-gluon corrections at NLO account for about 90\% of the total NLO corrections, thus proving that the soft-gluon contributions are overwhelmingly dominant numerically. To assess perturbative convergence, we examine $K$-factors defined as bin-by-bin ratios of differential cross sections at successive orders. We first discuss the total cross section, then the individual distributions at each order, and finally the distributions compared across orders together with the resulting $K$-factors.
 
\begin{table}[H]
\begin{center}
\begin{tabular}{|c|c|c|c|}
        \hline
        Order & Collider energy~[TeV] & PDF order & $\sigma$~[pb] \\
        \hline
        \multirow{4}{*}{LO}
        & \multirow{2}{*}{13}   & nnlo  & $46.7 \pm 4.1 \pm 0.6$ \\
        &    & an3lo & $47.0 \pm 4.0 \pm 0.6$ \\
        \cline{2-4}
        & \multirow{2}{*}{13.6} & nnlo  & $49.2 \pm 4.4 \pm 0.7$ \\
        &  & an3lo & $49.4 \pm 4.3 \pm 0.7$ \\
        \hline
        \multirow{4}{*}{NLO}
        & \multirow{2}{*}{13}   & nnlo  & $67.1 \pm 0.6 \pm 0.8$ \\
        &    & an3lo & $67.3 \pm 0.6 \pm 0.9$ \\
        \cline{2-4}
        & \multirow{2}{*}{13.6} & nnlo  & $70.7 \pm 0.6 \pm 0.8$ \\
        &  & an3lo & $71.0 \pm 0.6 \pm 0.9$ \\
        \hline
        \multirow{4}{*}{aNNLO}
        & \multirow{2}{*}{13}   & nnlo  & $73.2 \pm 1.4 \pm 0.9$ \\
        &    & an3lo & $73.5 \pm 1.4 \pm 0.9$ \\
        \cline{2-4}
        & \multirow{2}{*}{13.6} & nnlo  & $77.1 \pm 1.5 \pm 0.9$ \\
        &  & an3lo & $77.5 \pm 1.5 \pm 1.0$ \\
        \hline
        \multirow{4}{*}{aN$^3$LO}
        & \multirow{2}{*}{13}   & nnlo  & $74.4 \pm 2.6 \pm 0.9$ \\
        &    & an3lo & $74.7 \pm 2.6 \pm 1.0$ \\
        \cline{2-4}
        & \multirow{2}{*}{13.6} & nnlo  & $78.4 \pm 2.7 \pm 0.9$ \\
        &  & an3lo & $78.8 \pm 2.8 \pm 1.0$ \\
        \hline
\end{tabular}
\caption{Integrated fiducial cross sections for $pp \to \gamma Z$ production at the LHC with $p_T^\gamma > 20$ GeV using MSHT20 PDFs. The first uncertainty is the 3-point scale variation and the second is the PDF uncertainty.}
\label{tab:total_xsec}
\end{center}
\end{table}

Table~\ref{tab:total_xsec} presents integrated fiducial cross sections for $pp \to \gamma Z$ production with $p_T^\gamma > 20$ GeV at 13 TeV and 13.6 TeV energies computed at LO, NLO QCD, aNNLO QCD, and aN$^3$LO QCD using MSHT20 nnlo and an3lo PDF sets. 

Scale uncertainties decrease substantially with perturbative order, from roughly 9\% at LO to below 1\% at NLO, increasing again to approximately 2\% at aNNLO and 3.5\% at aN$^3$LO due to the quadrature addition of scale uncertainties from each successive soft-gluon contribution. PDF uncertainties remain stable at 1.2\%--1.3\% across all orders, with an3lo PDFs yielding slightly larger variations than nnlo. The NLO corrections enhance the LO prediction by a $K$-factor of approximately 1.44, while the aNNLO result exceeds the NLO prediction by roughly 9\%, and the aN$^3$LO result is larger than aNNLO by about 2\%, indicating good perturbative convergence. Cross sections increase by approximately 5\%--6\% from 13 to 13.6~TeV at all perturbative orders. The choice between nnlo and an3lo PDFs has minimal impact on the central values, with differences well within the quoted PDF uncertainties.

\begin{figure}
    [H]
    \centering
    \includegraphics[width=0.4875\linewidth]{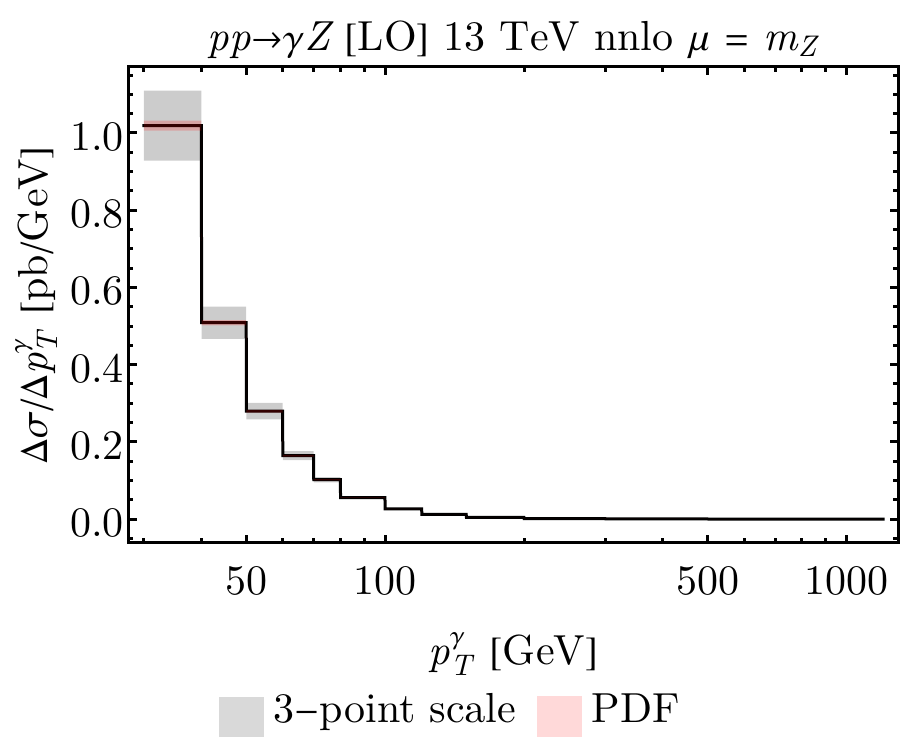}
    \includegraphics[width=0.4875\linewidth]{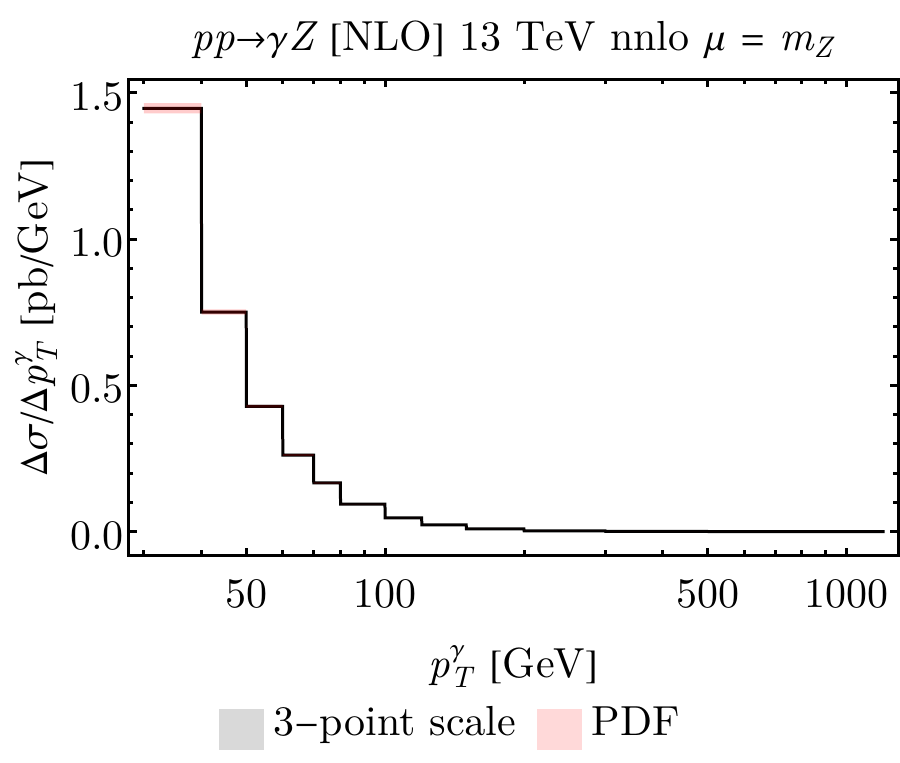}
    \includegraphics[width=0.4875\linewidth]{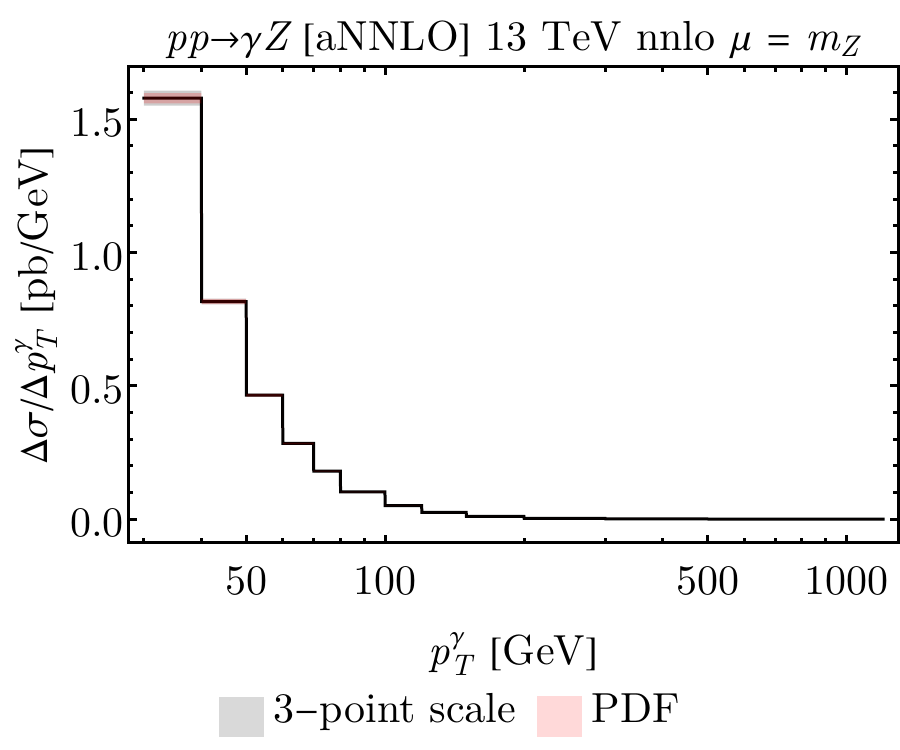}
    \includegraphics[width=0.4875\linewidth]{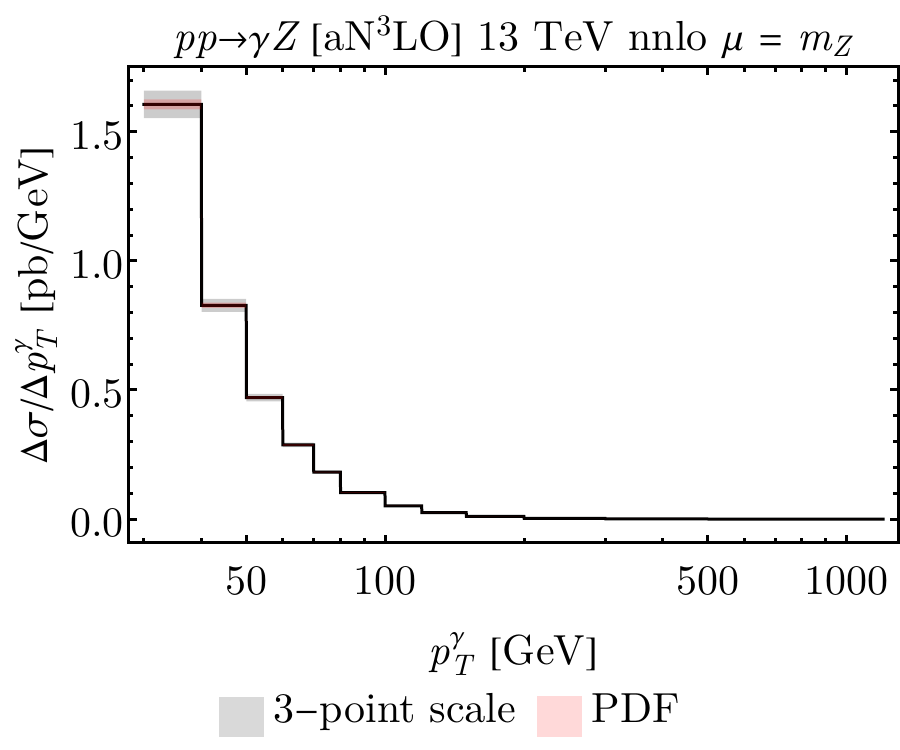}
    \caption{Single-differential $p_T^\gamma$ distributions at 13~TeV with MSHT20 nnlo PDFs, showing scale and PDF uncertainties, at LO (top left), NLO QCD (top right), aNNLO QCD (bottom left), and aN$^3$LO QCD (bottom right).}
    \label{fig:individual_distributions_13_nnlo_pt}
\end{figure}

\begin{figure}
    [H]
    \centering
    \includegraphics[width=0.4875\linewidth]{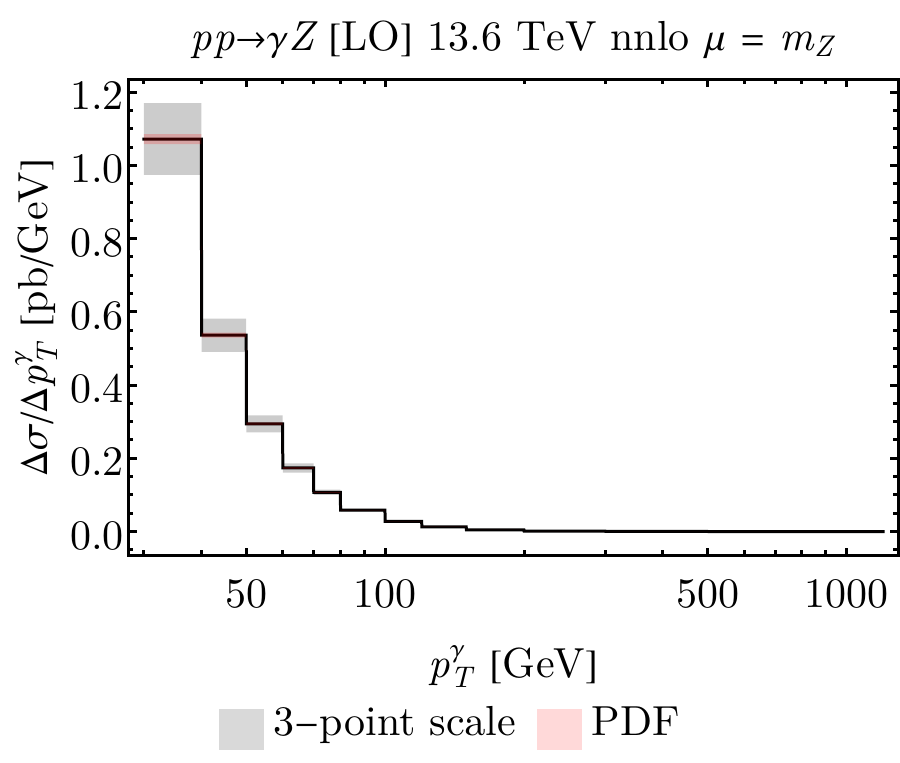}
    \includegraphics[width=0.4875\linewidth]{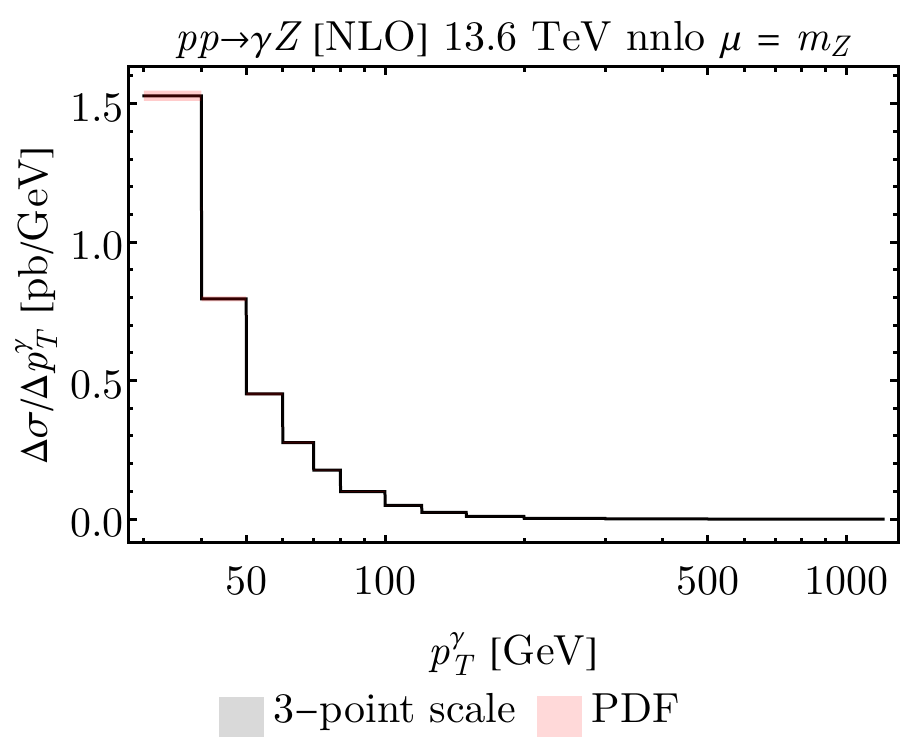}
    \includegraphics[width=0.4875\linewidth]{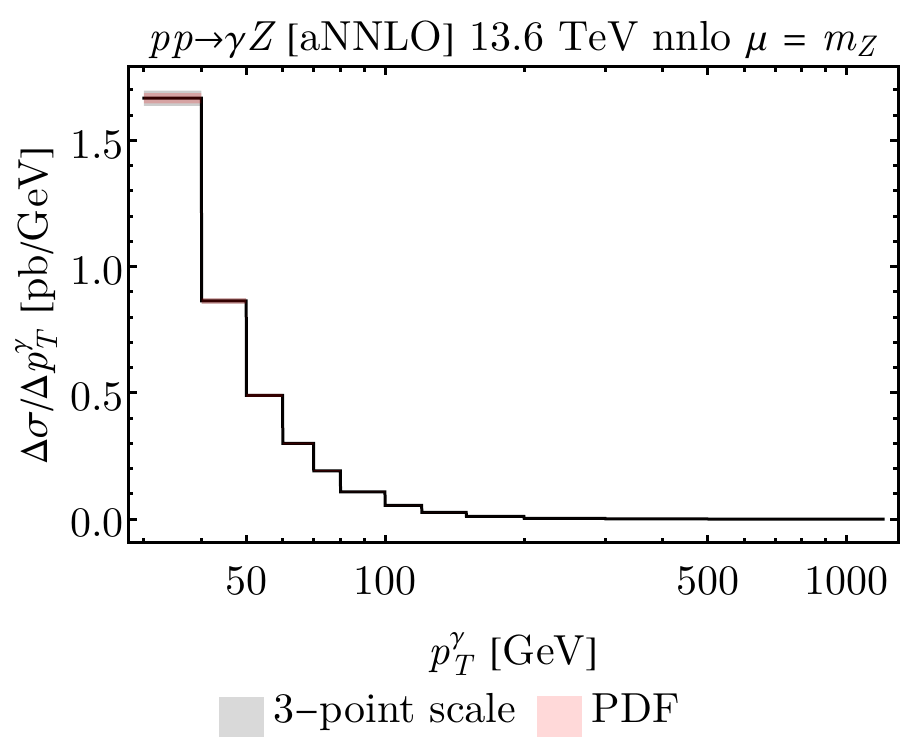}
    \includegraphics[width=0.4875\linewidth]{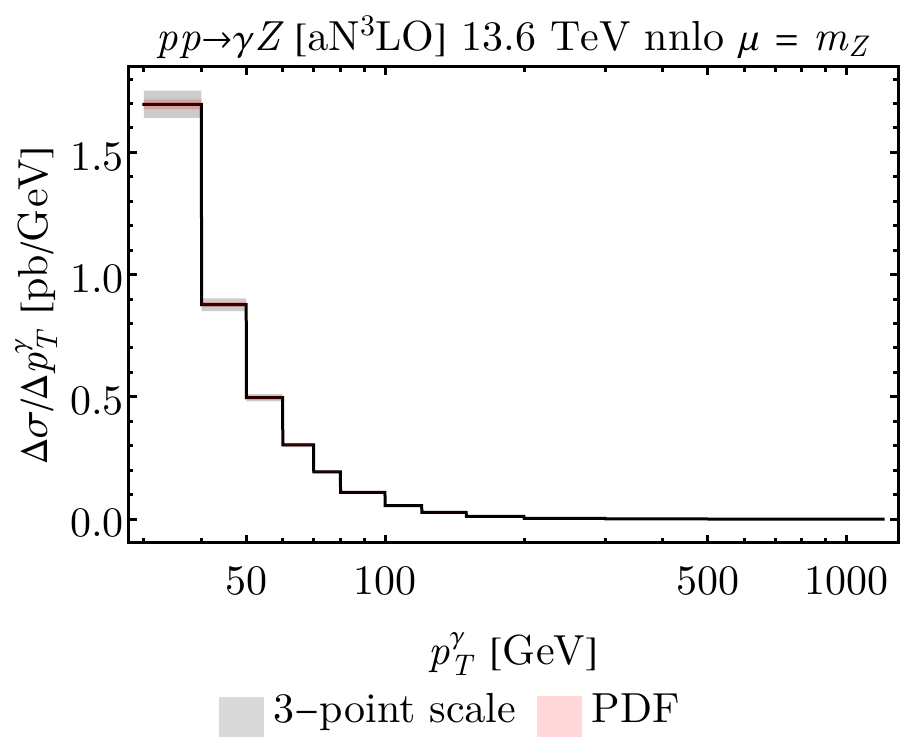}
    \caption{The same as Figure~\ref{fig:individual_distributions_13_nnlo_pt} but at 13.6~TeV.}
    \label{fig:individual_distributions_13.6_nnlo_pt}
\end{figure}

\begin{figure}
    [H]
    \centering
    \includegraphics[width=0.4875\linewidth]{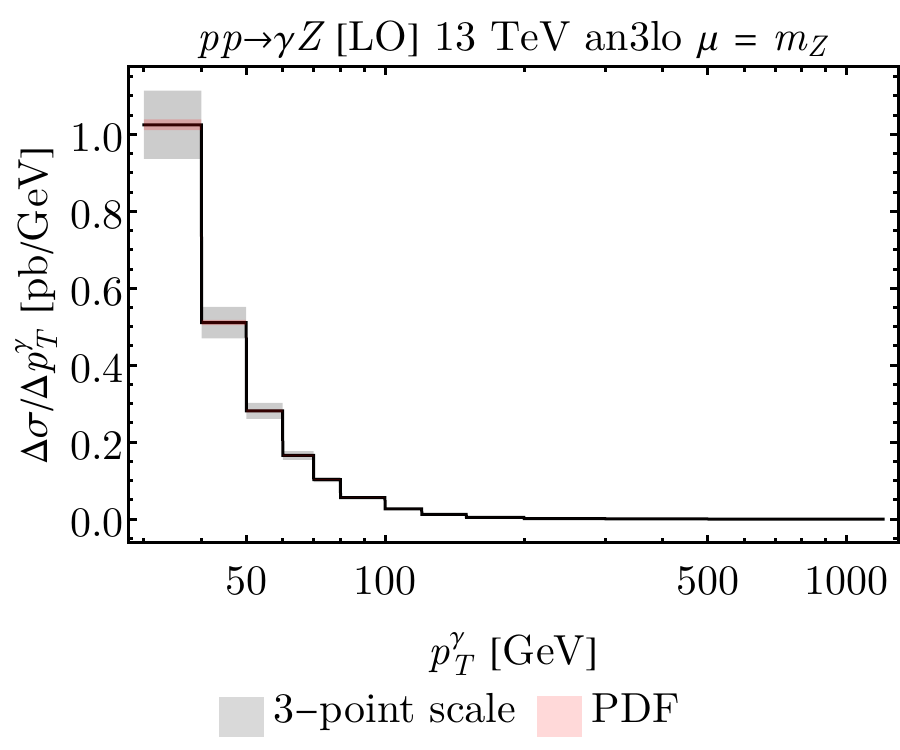}
    \includegraphics[width=0.4875\linewidth]{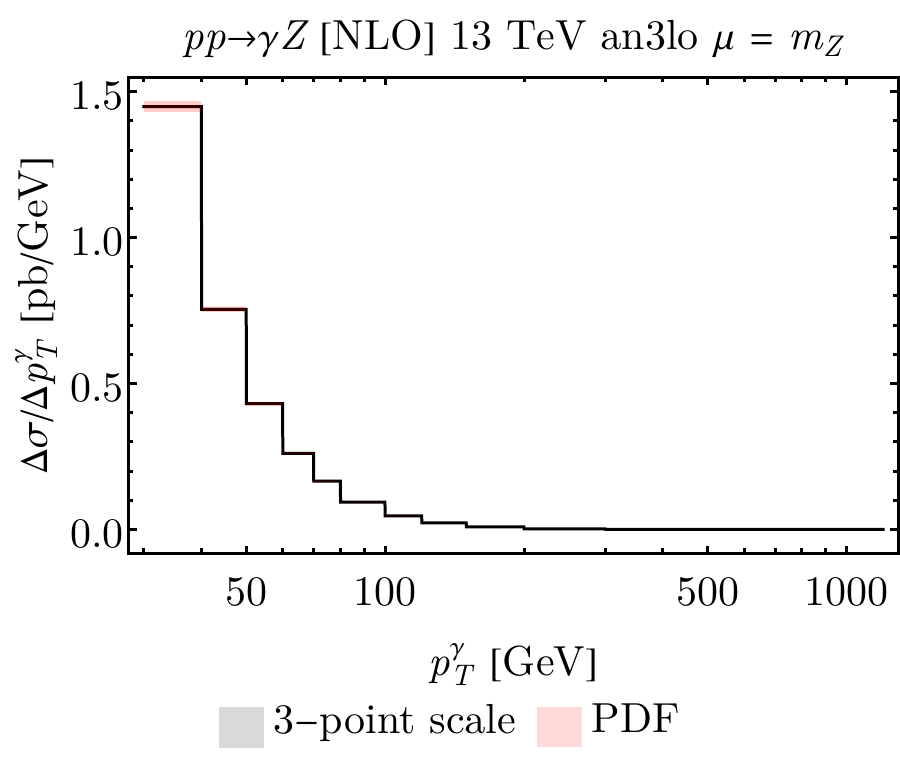}
    \includegraphics[width=0.4875\linewidth]{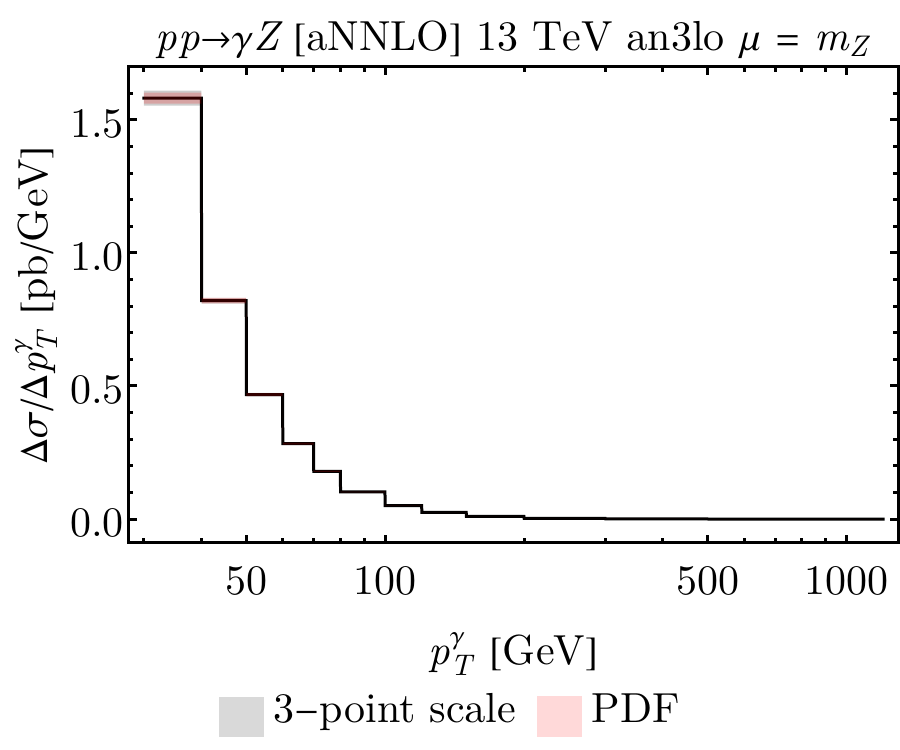}
    \includegraphics[width=0.4875\linewidth]{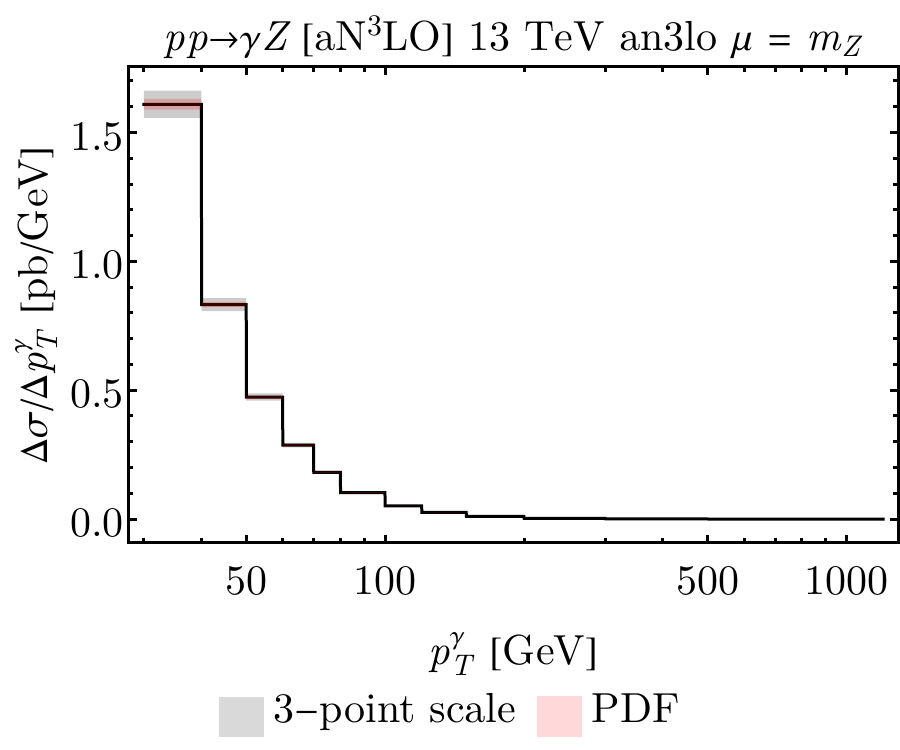}
    \caption{The same as Figure~\ref{fig:individual_distributions_13_nnlo_pt} but with MSHT20 an3lo PDFs.}
    \label{fig:individual_distributions_13_an3lo_pt}
\end{figure}

\begin{figure}
    [H]
    \centering
    \includegraphics[width=0.4875\linewidth]{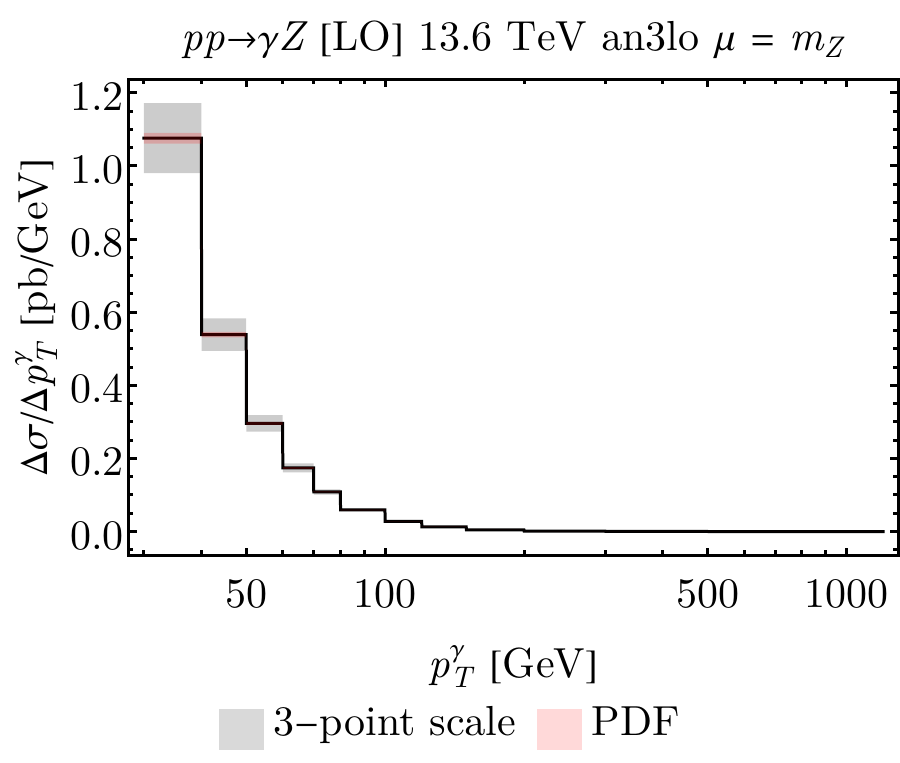}
    \includegraphics[width=0.4875\linewidth]{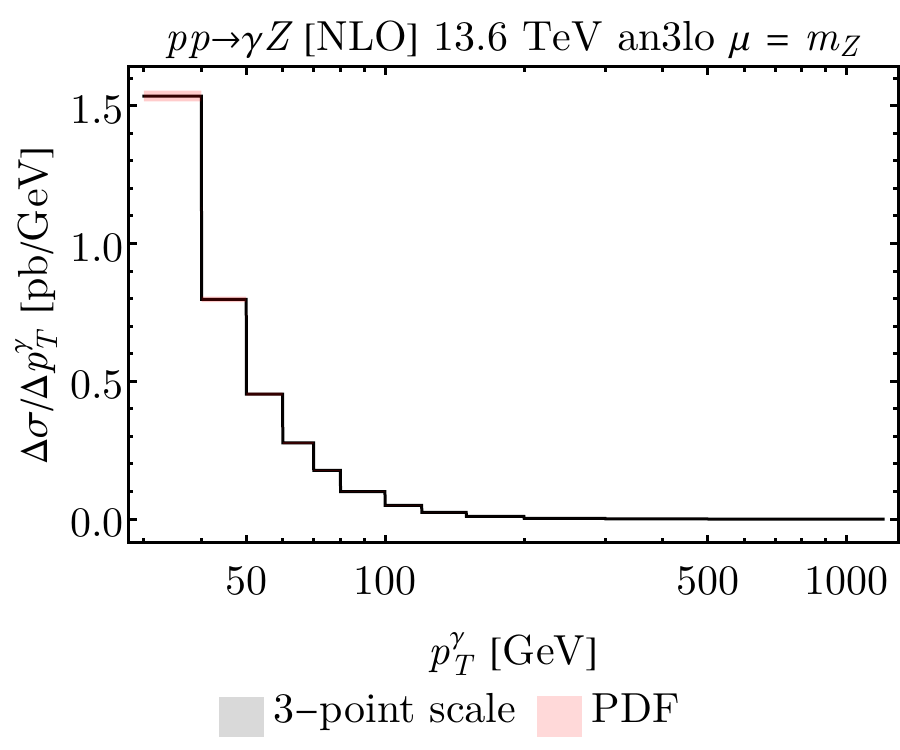}
    \includegraphics[width=0.4875\linewidth]{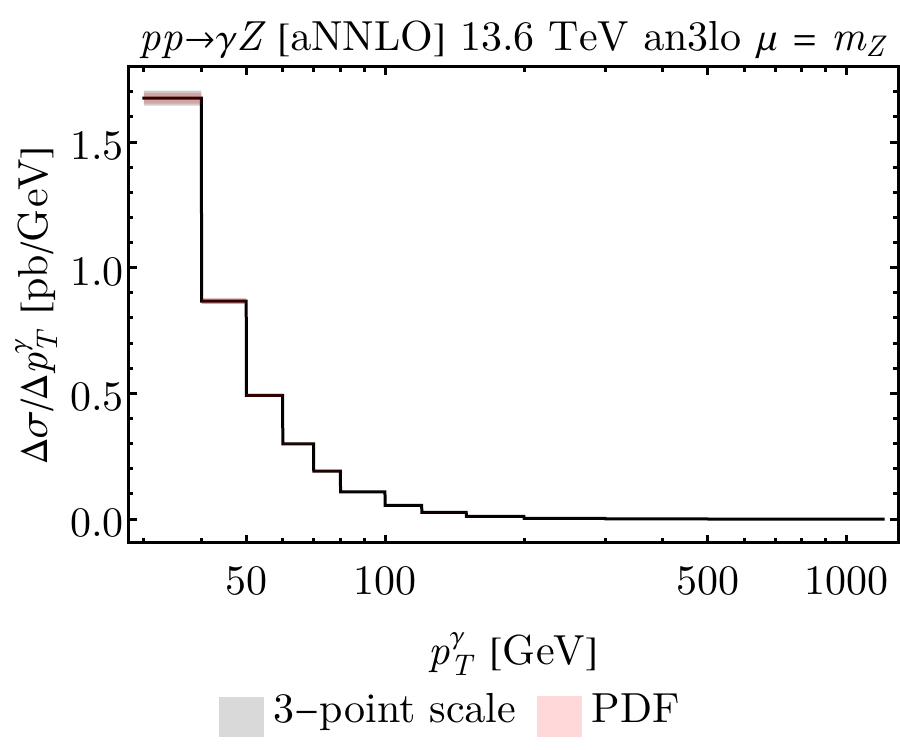}
    \includegraphics[width=0.4875\linewidth]{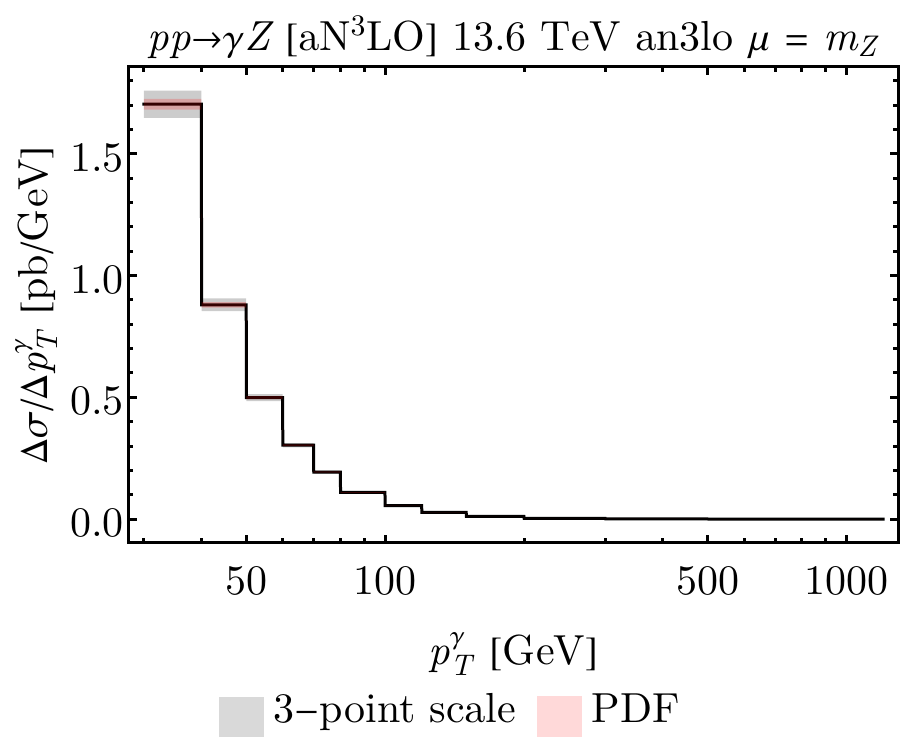}
    \caption{The same as Figure~\ref{fig:individual_distributions_13_an3lo_pt} but at 13.6~TeV.}
    \label{fig:individual_distributions_13.6_an3lo_pt}
\end{figure}

\begin{figure}
    [H]
    \centering
    \includegraphics[width=0.4875\linewidth]{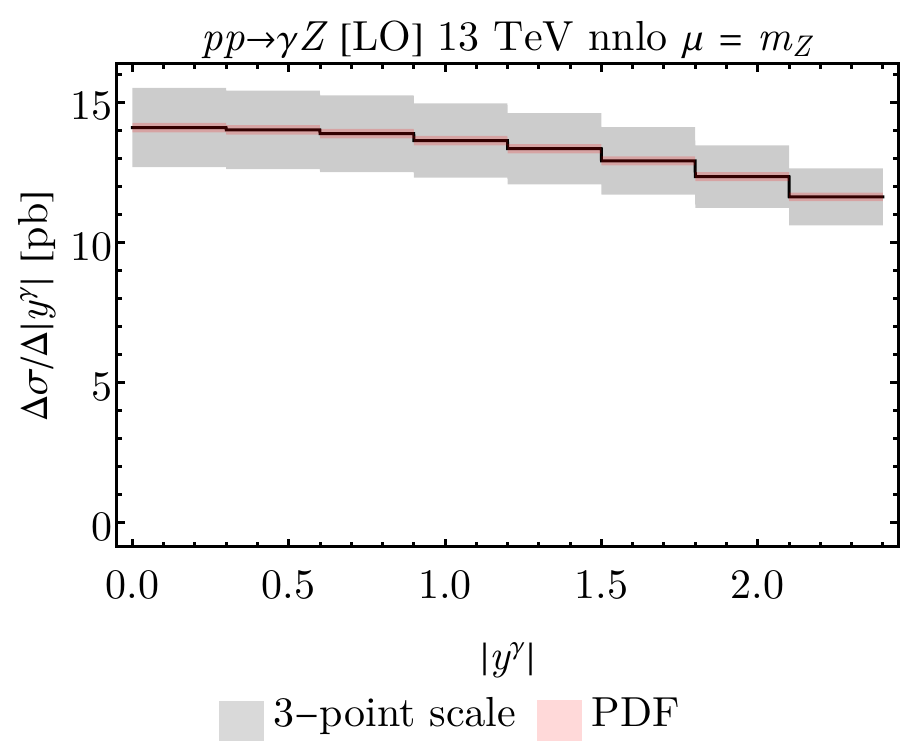}
    \includegraphics[width=0.4875\linewidth]{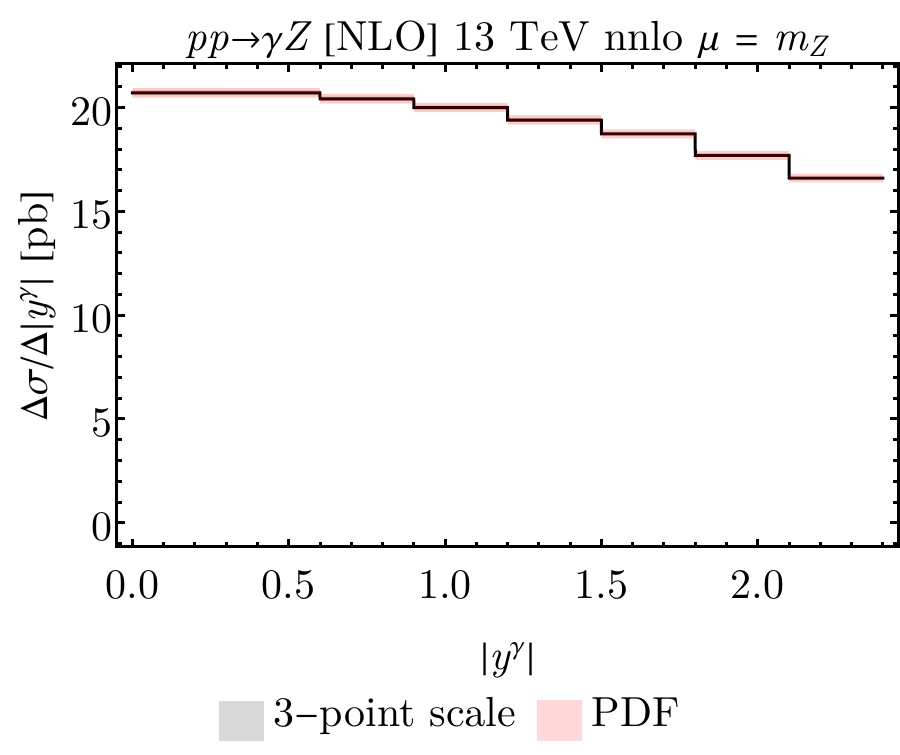}
    \includegraphics[width=0.4875\linewidth]{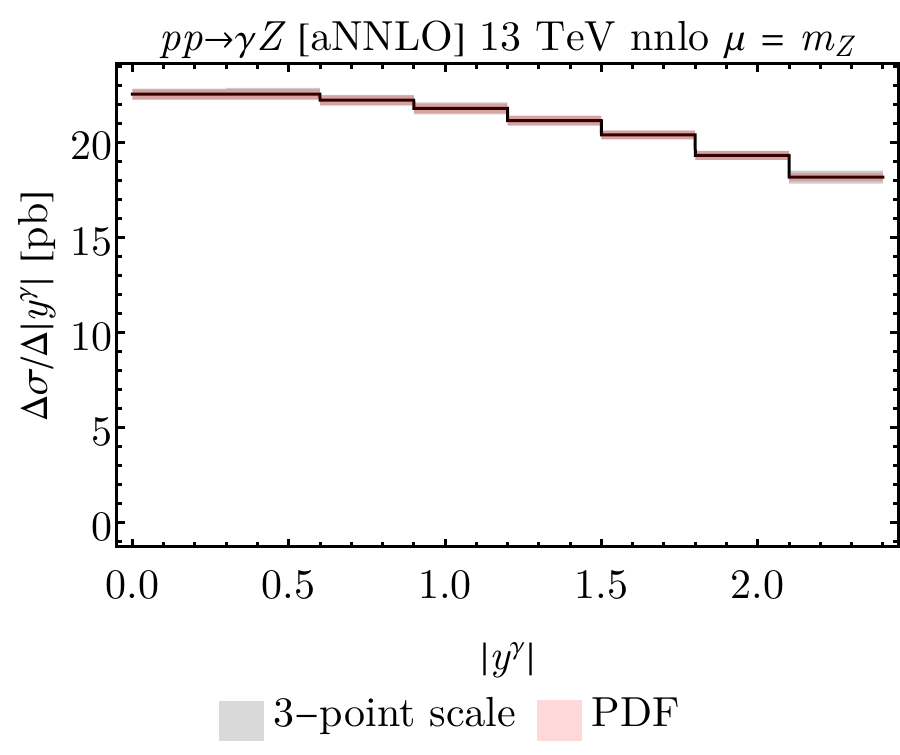}
    \includegraphics[width=0.4875\linewidth]{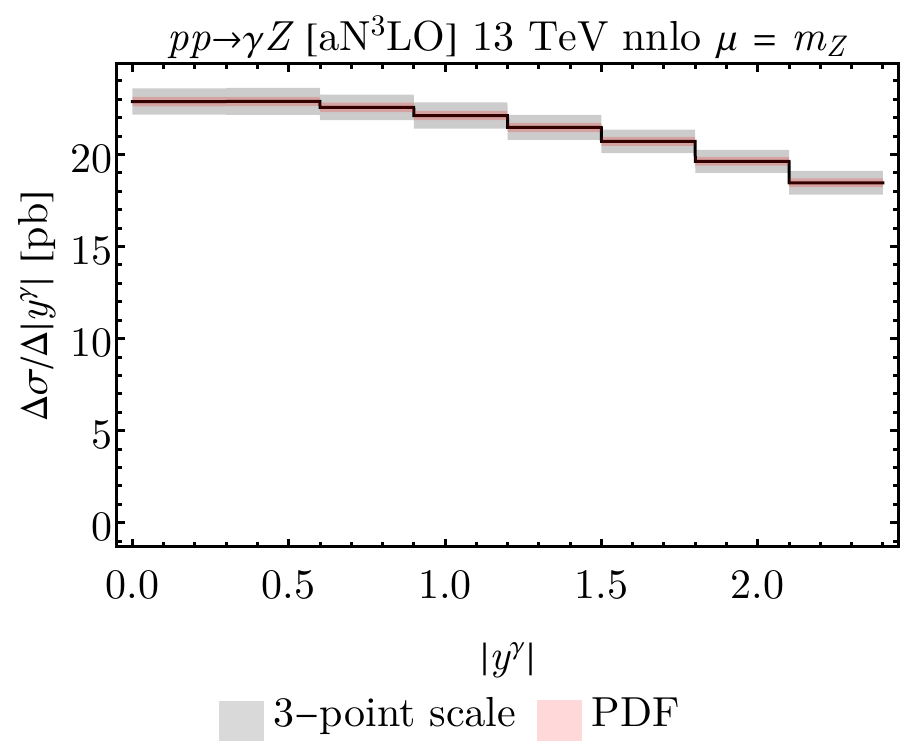}
    \caption{Single-differential $|y^\gamma|$ distributions at 13~TeV with MSHT20 nnlo PDFs, showing scale and PDF uncertainties, at LO (top left), NLO QCD (top right), aNNLO QCD (bottom left), and aN$^3$LO QCD (bottom right).}
    \label{fig:individual_distributions_13_nnlo_absy}
\end{figure}

\begin{figure}
    [H]
    \centering
    \includegraphics[width=0.4875\linewidth]{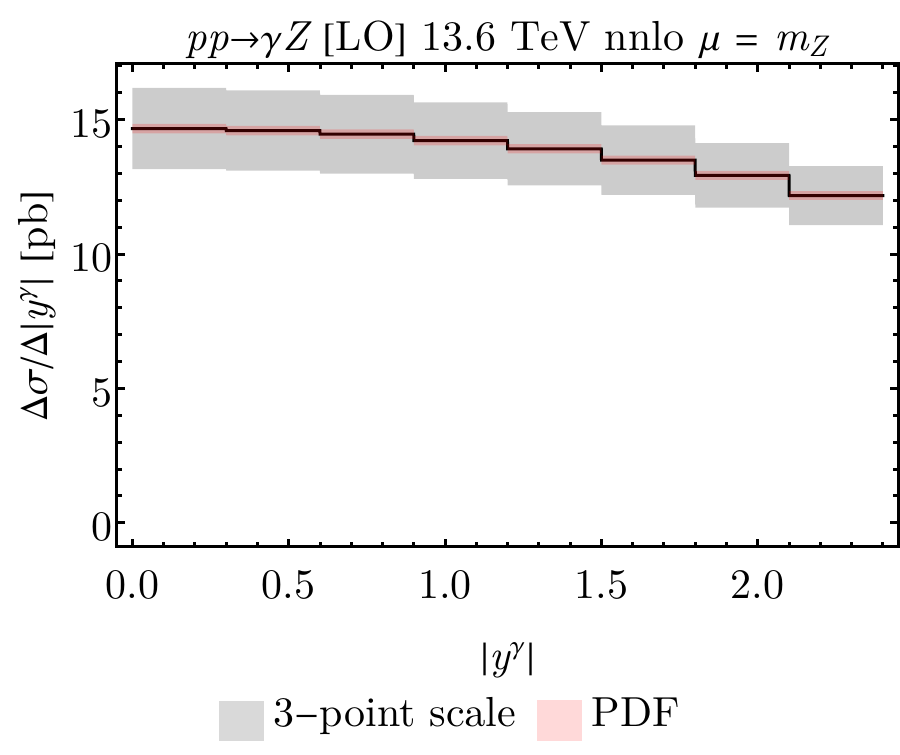}
    \includegraphics[width=0.4875\linewidth]{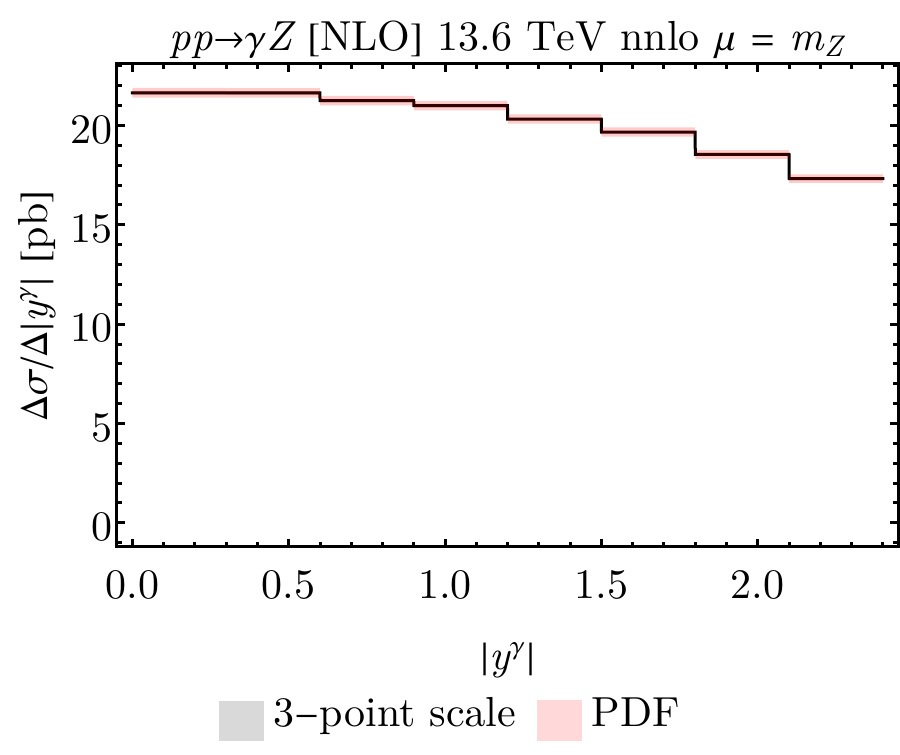}
    \includegraphics[width=0.4875\linewidth]{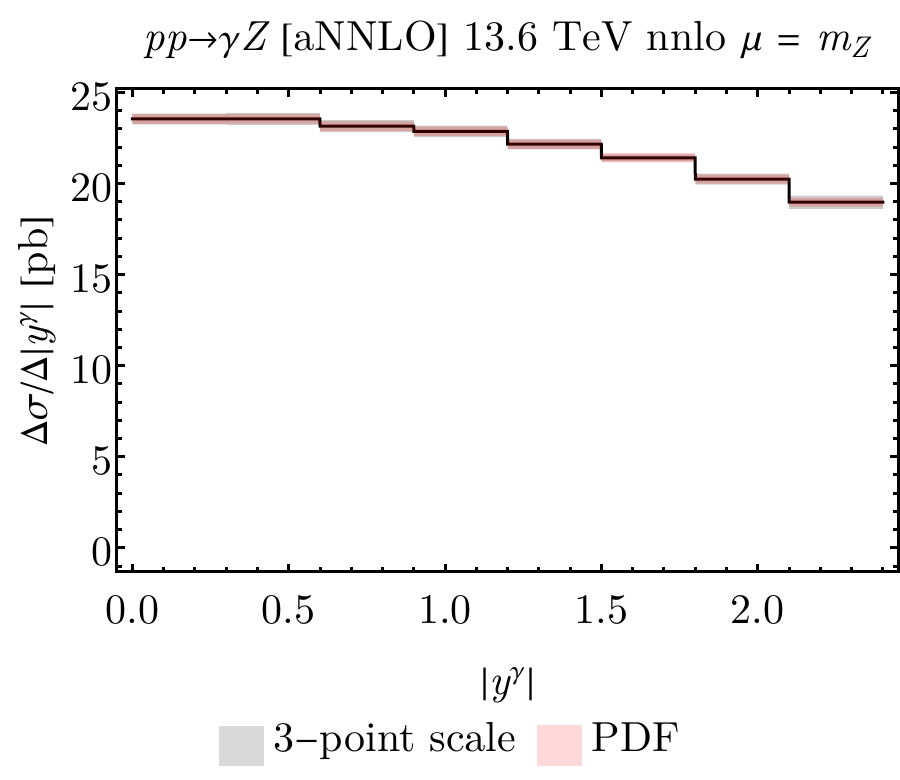}
    \includegraphics[width=0.4875\linewidth]{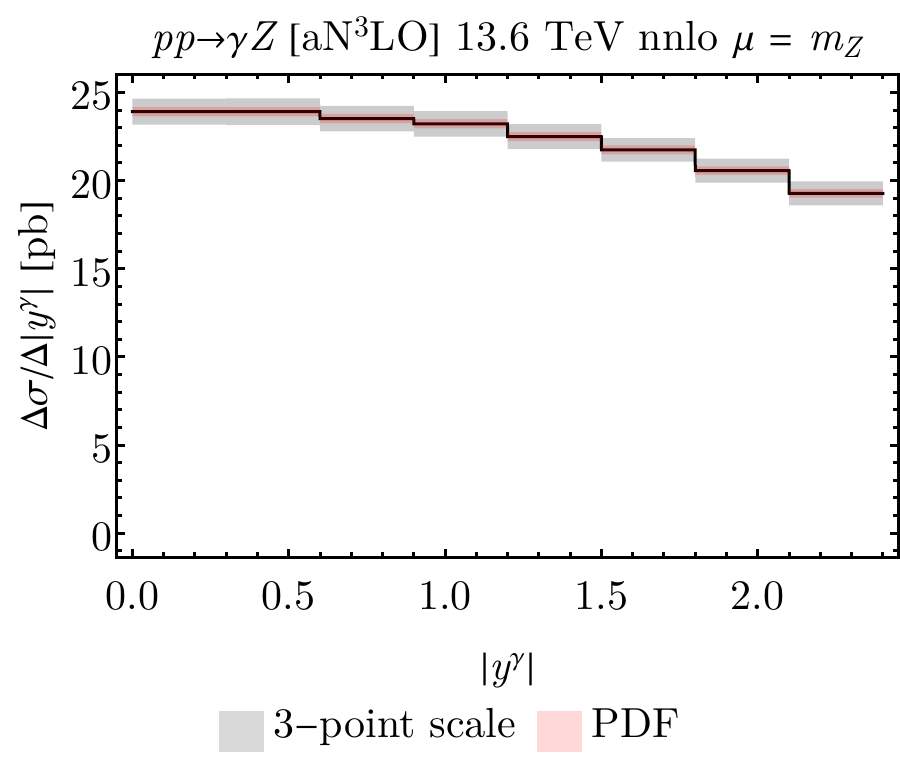}
    \caption{The same as Figure~\ref{fig:individual_distributions_13_nnlo_absy} but at 13.6~TeV.}
    \label{fig:individual_distributions_13.6_nnlo_absy}
\end{figure}

\begin{figure}
    [H]
    \centering
    \includegraphics[width=0.4875\linewidth]{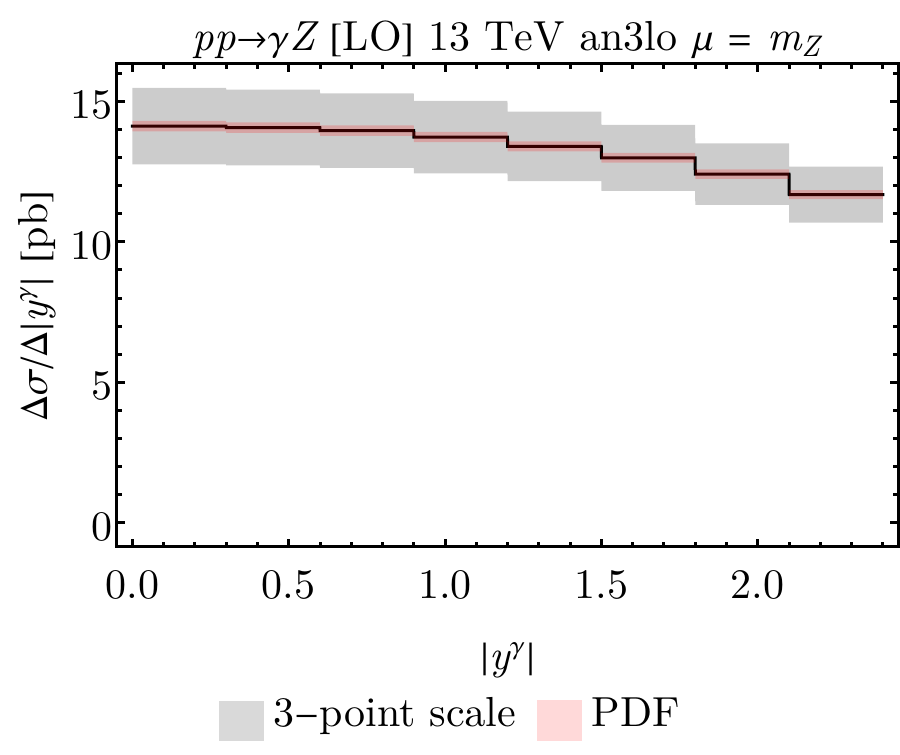}
    \includegraphics[width=0.4875\linewidth]{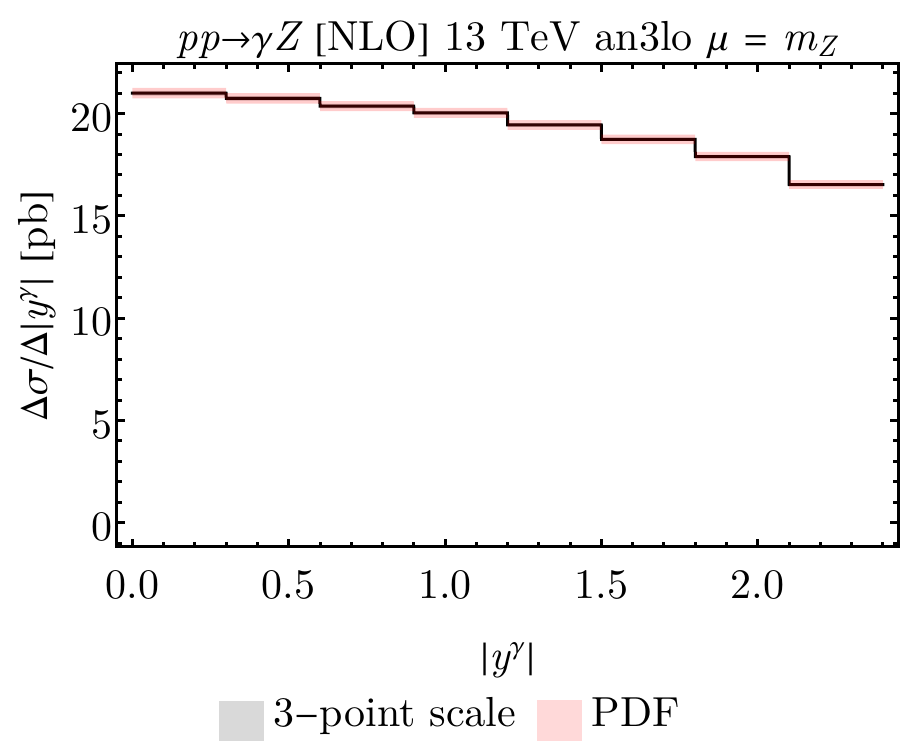}
    \includegraphics[width=0.4875\linewidth]{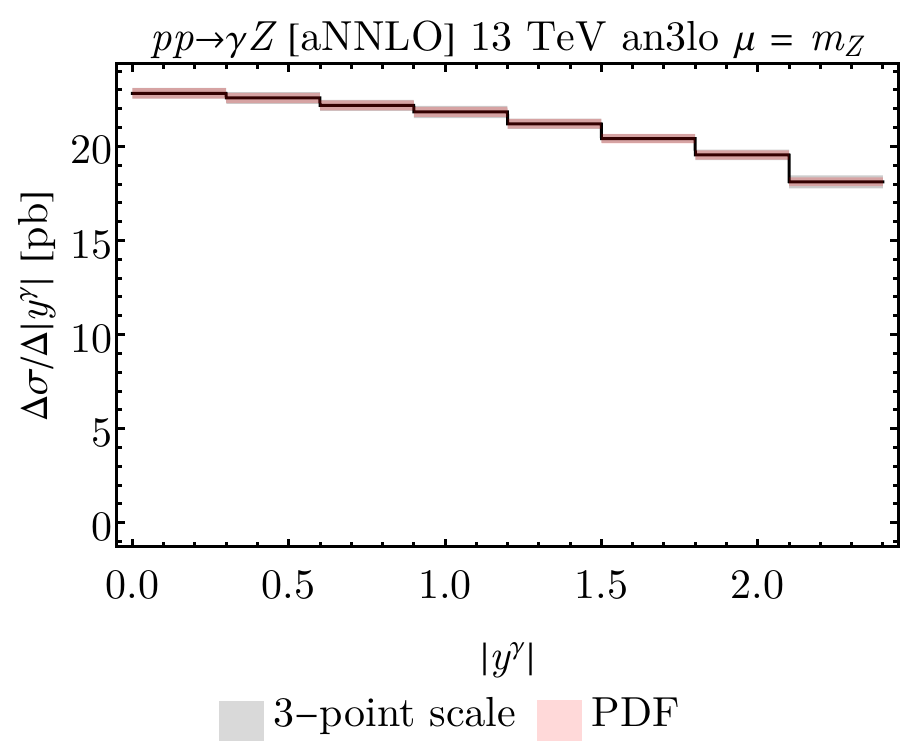}
    \includegraphics[width=0.4875\linewidth]{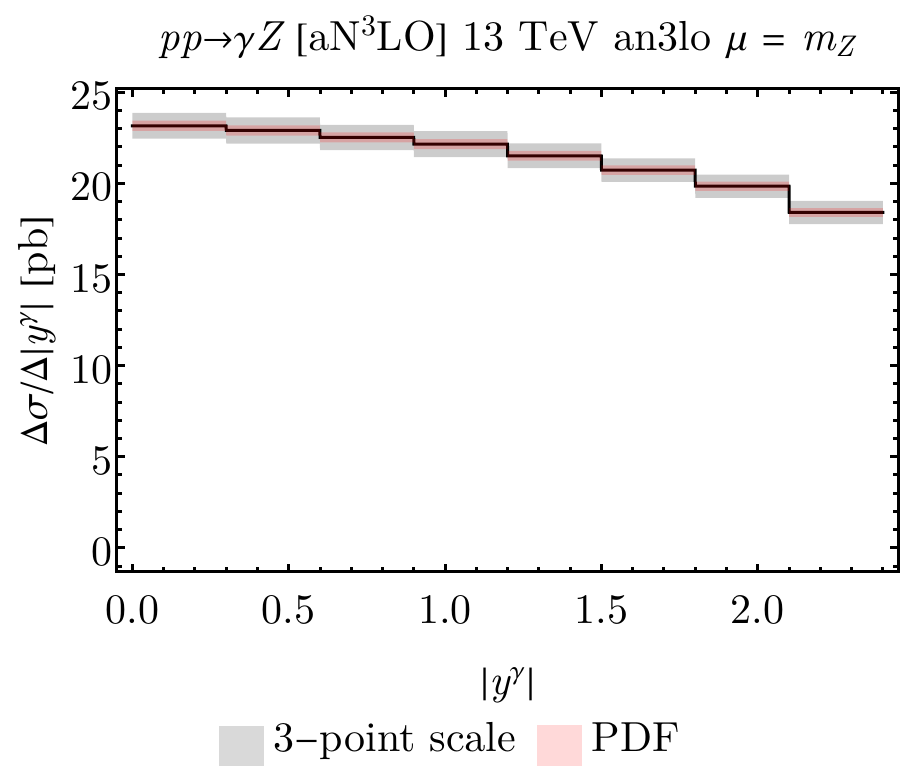}
    \caption{The same as Figure~\ref{fig:individual_distributions_13_nnlo_absy} but with MSHT20 an3lo PDFs.}
    \label{fig:individual_distributions_13_an3lo_absy}
\end{figure}

\begin{figure}
    [H]
    \centering
    \includegraphics[width=0.4875\linewidth]{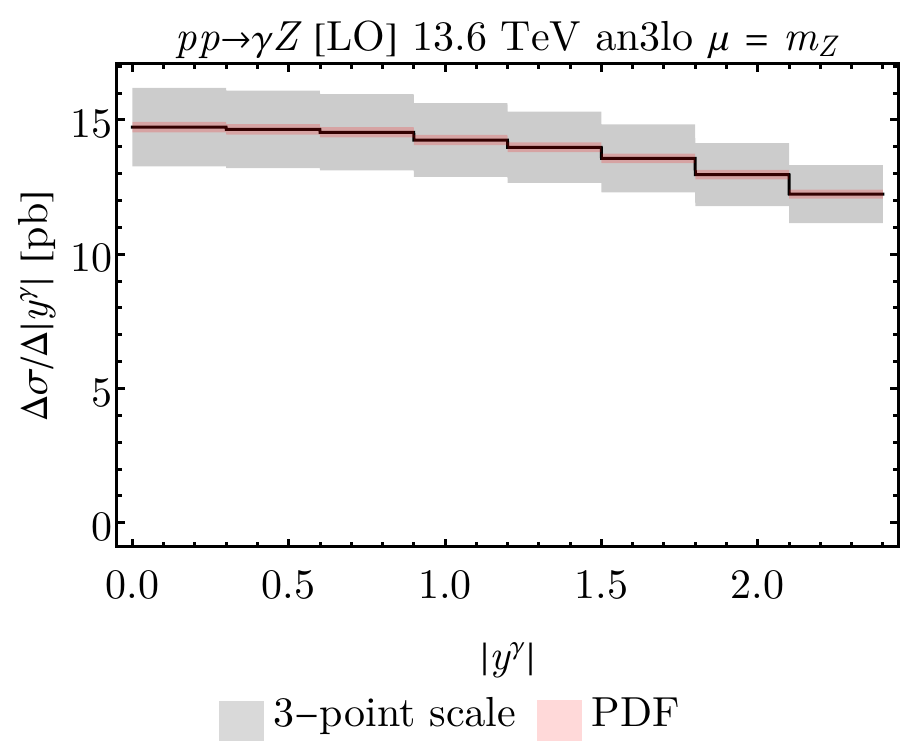}
    \includegraphics[width=0.4875\linewidth]{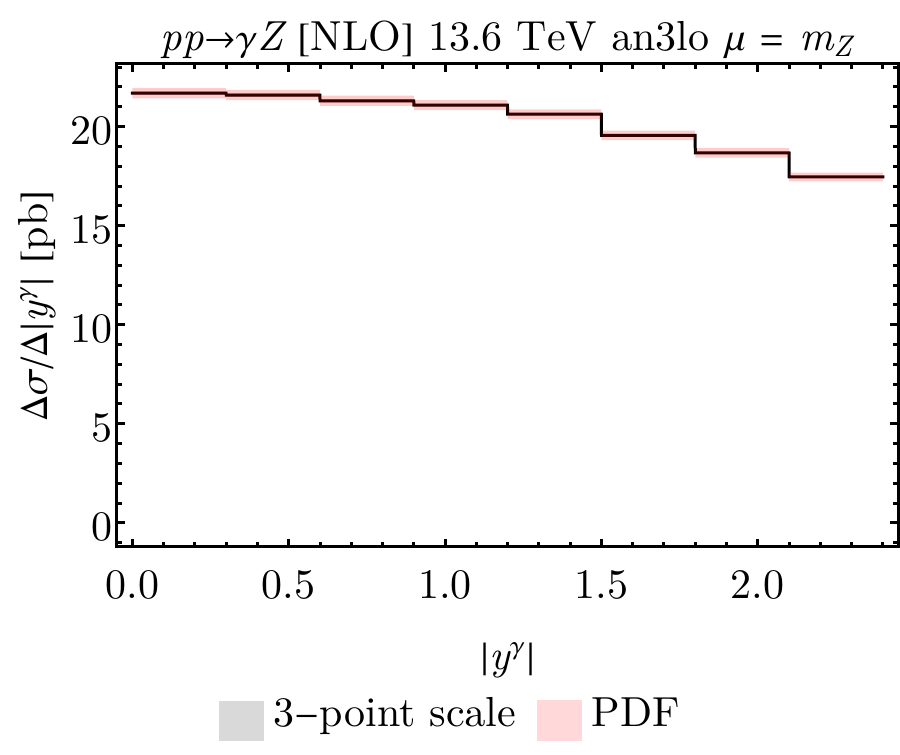}
    \includegraphics[width=0.4875\linewidth]{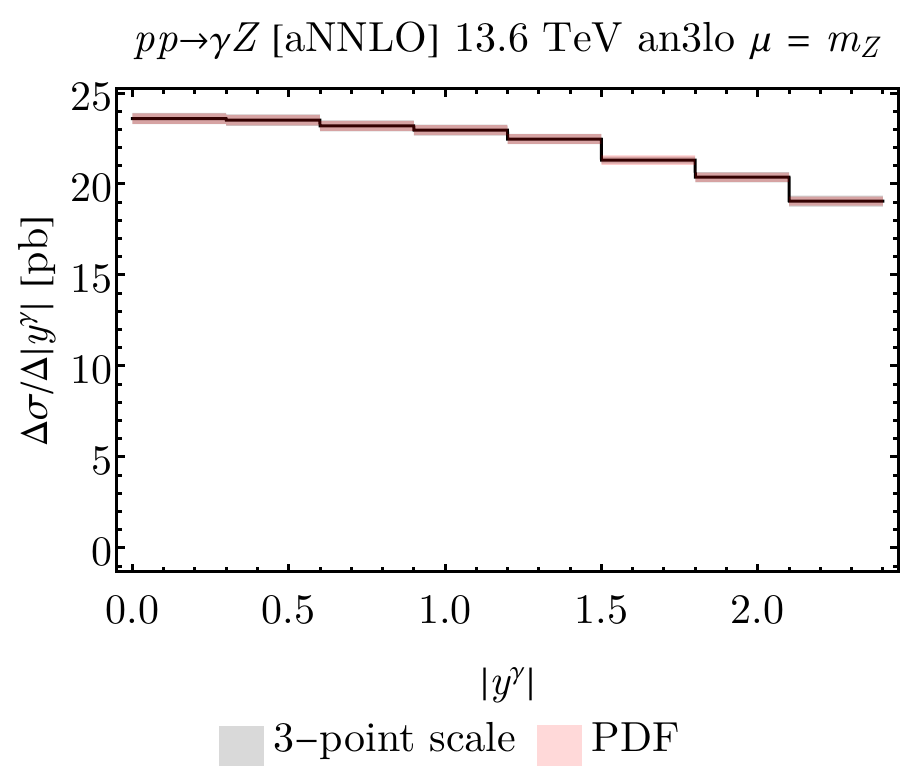}
    \includegraphics[width=0.4875\linewidth]{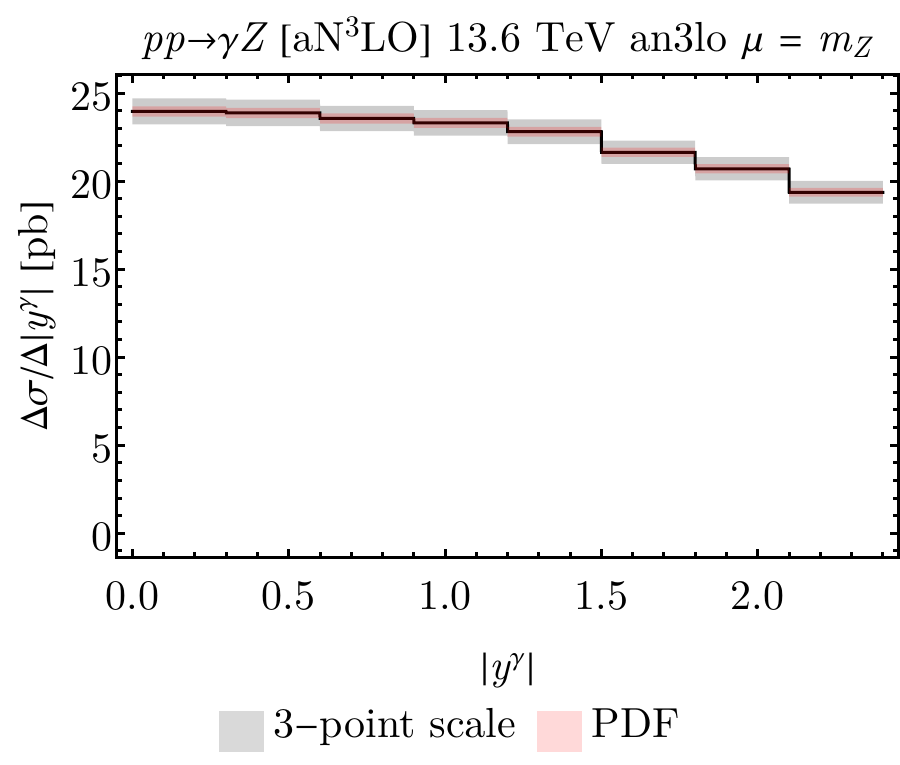}
    \caption{The same as Figure~\ref{fig:individual_distributions_13_an3lo_absy} but at 13.6~TeV.}
    \label{fig:individual_distributions_13.6_an3lo_absy}
\end{figure}

Figures~\ref{fig:individual_distributions_13_nnlo_pt}--\ref{fig:individual_distributions_13.6_an3lo_absy} present single differential $p_T^\gamma$ and $|y^\gamma|$ distributions at 13 and 13.6~TeV for both MSHT20 nnlo and an3lo PDF sets. Each figure displays four panels corresponding to LO (top left), NLO (top right), aNNLO (bottom left), and aN$^3$LO (bottom right) predictions. The solid black curves represent the central predictions, while the gray bands indicate 3-point scale variations obtained by varying the renormalization and factorization scales by factors of 2 and 0.5 around the central choice $\mu_R = \mu_F = m_Z$. The red bands show PDF uncertainties computed using the Hessian method with the respective MSHT20 error sets. The $p_T^\gamma$ distributions exhibit at least a full order of magnitude variation across the kinematic range, whereas the $|y^\gamma|$ distributions are comparatively flat, with both observables showing consistent behavior across the two collider energies. 


Scale uncertainties in the $p_T^\gamma$ distribution span a wide range across bins within each order: at LO the per-bin uncertainty ranges from roughly 2\% to 9\%, decreasing to below 0.2\% at low $p_T^\gamma$ and reaching up to 16\% in the high-$p_T^\gamma$ tail at NLO QCD, and increasing further to up to 26\%--27\% at aNNLO QCD and aN$^3$LO QCD, where the quadrature addition of soft-gluon scale uncertainties dominates. For $|y^\gamma|$, the uncertainties are more uniform across bins: at LO they range from 8.7\% to 10\% and decrease to below 0.6\% at NLO QCD. At aNNLO QCD the $|y^\gamma|$ uncertainties increase to 0.9\%--1.9\%, and at aN$^3$LO QCD they increase further to 3.0\%--3.5\%, reflecting the quadrature addition of scale uncertainties from each successive soft-gluon contribution, since aNNLO and aN$^3$LO are constructed by adding the approximate higher-order corrections to the NLO result. The PDF uncertainty ranges are stable across both PDF orders and both collider energies, with variations of at most 0.3\% between nnlo and an3lo PDFs and at most 0.4\% between 13 and 13.6~TeV. The PDF uncertainties in the $p_T^\gamma$ distribution show a mild increase toward high $p_T^\gamma$, reaching approximately 2.7\%--2.9\% in the highest bins, reflecting the larger uncertainties on large-$x$ PDFs, while the $|y^\gamma|$ distribution exhibits uniform PDF uncertainties of 1.1\%--1.4\% across all bins.

\begin{figure}
    [H]
    \centering
    \includegraphics[width=0.4875\linewidth]{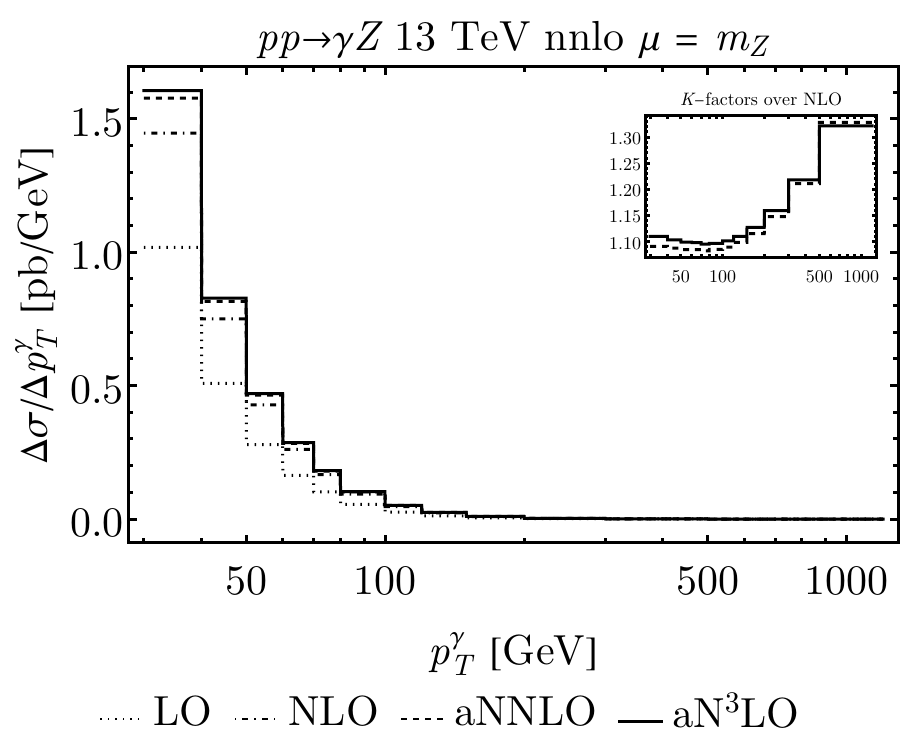}
    \includegraphics[width=0.4875\linewidth]{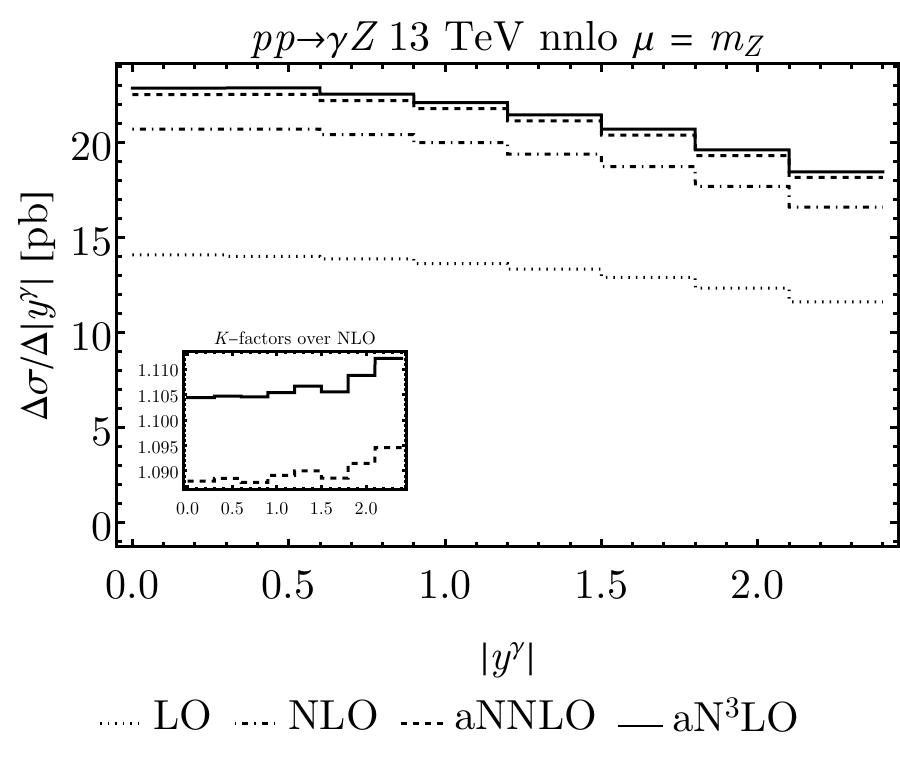}
    \caption{The $p_T^\gamma$ (left) and $|y^\gamma|$ (right) distributions at 13~TeV with MSHT20 nnlo PDFs at LO (dotted), NLO QCD (dot-dashed), aNNLO QCD (dashed), and aN$^3$LO QCD (solid). The inset shows the $K$-factors relative to NLO QCD, with aNNLO QCD (dashed) and aN$^3$LO QCD (solid).}
    \label{fig:comparative_distributions_13_nnlo}
\end{figure}

\begin{figure}
    [H]
    \centering
    \includegraphics[width=0.4875\linewidth]{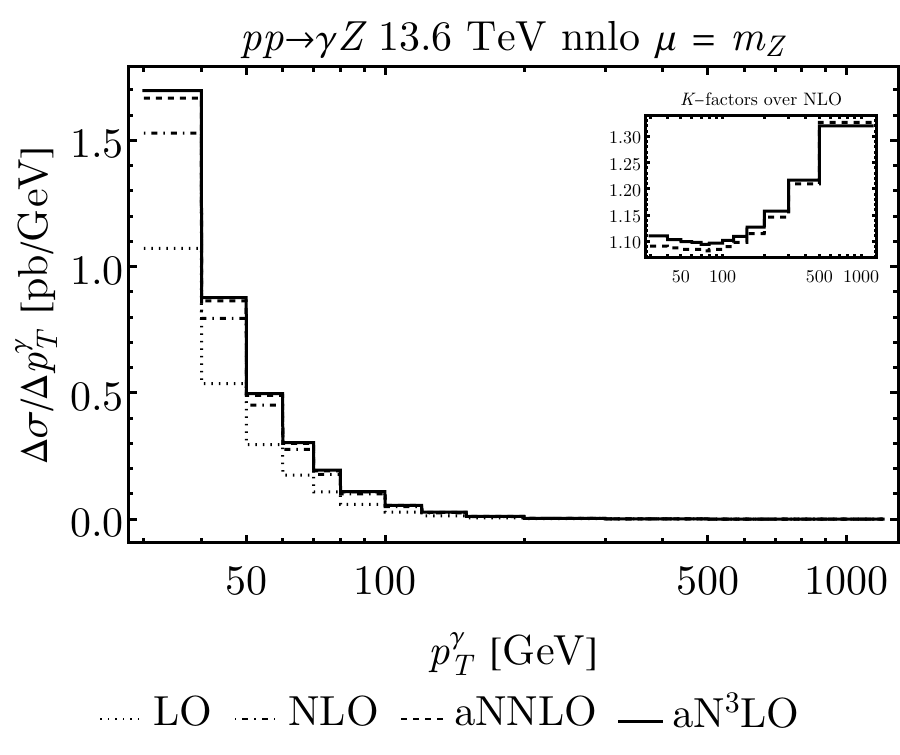}
    \includegraphics[width=0.4875\linewidth]{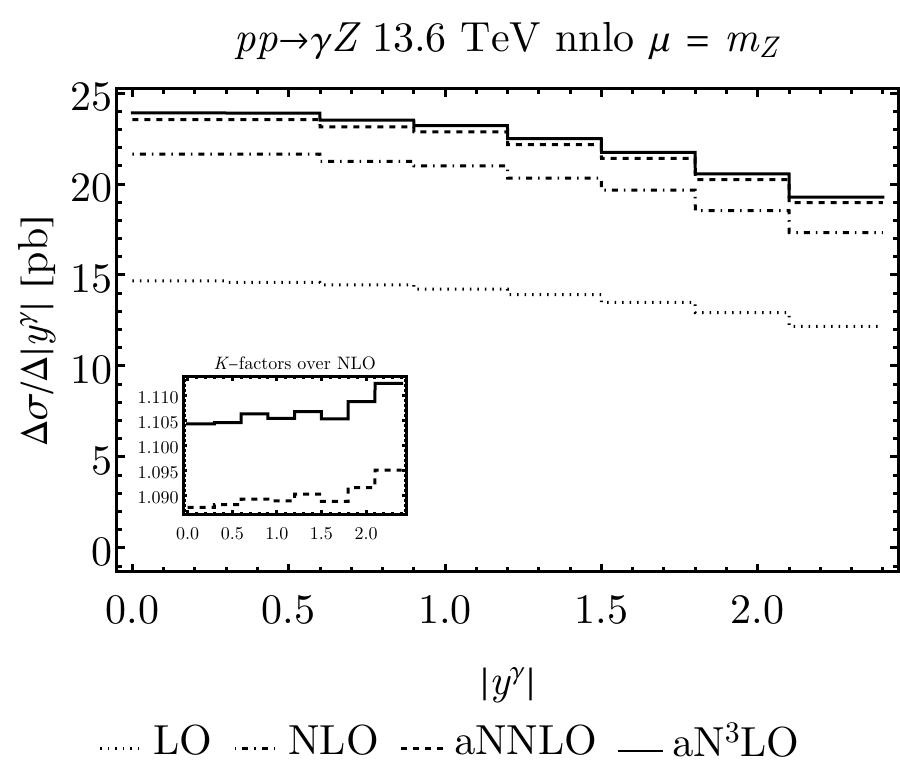}
    \caption{The same as Figure~\ref{fig:comparative_distributions_13_nnlo} but at 13.6~TeV.}
    \label{fig:comparative_distributions_13.6_nnlo}
\end{figure}

\begin{figure}
    [H]
    \centering
    \includegraphics[width=0.4875\linewidth]{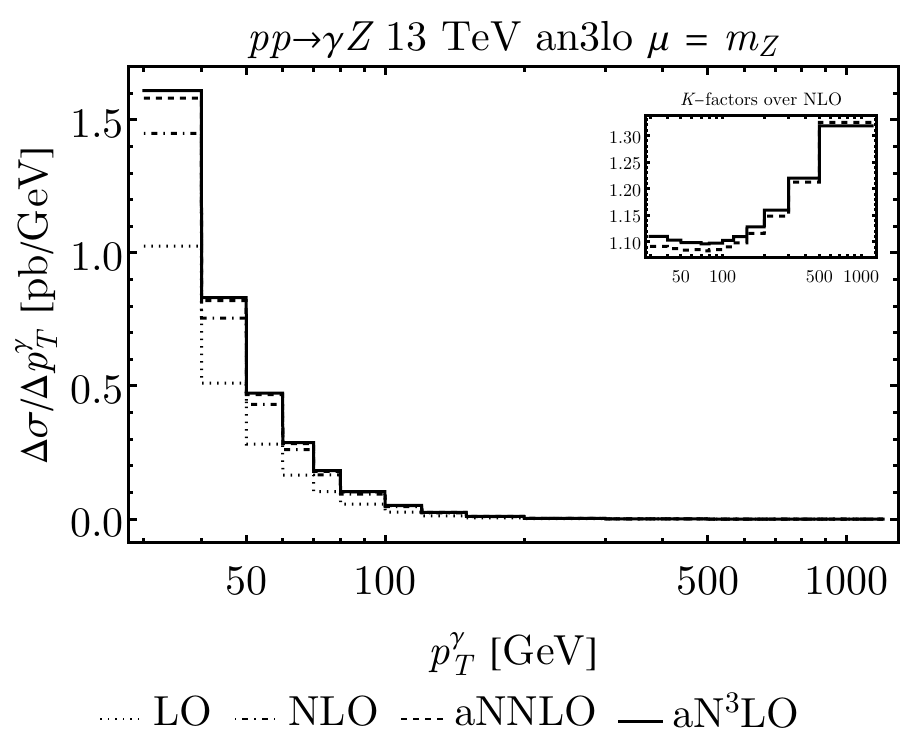}
    \includegraphics[width=0.4875\linewidth]{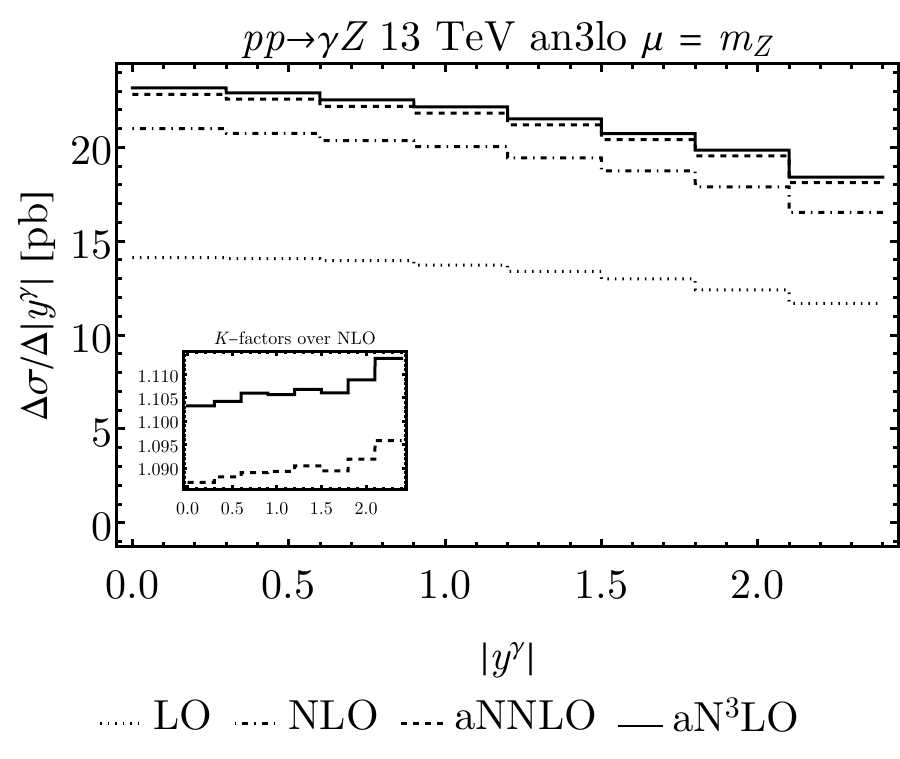}
    \caption{The same as Figure~\ref{fig:comparative_distributions_13_nnlo} but with MSHT20 an3lo PDFs.}
    \label{fig:comparative_distributions_13_an3lo}
\end{figure}

\begin{figure}
    [H]
    \centering
    \includegraphics[width=0.4875\linewidth]{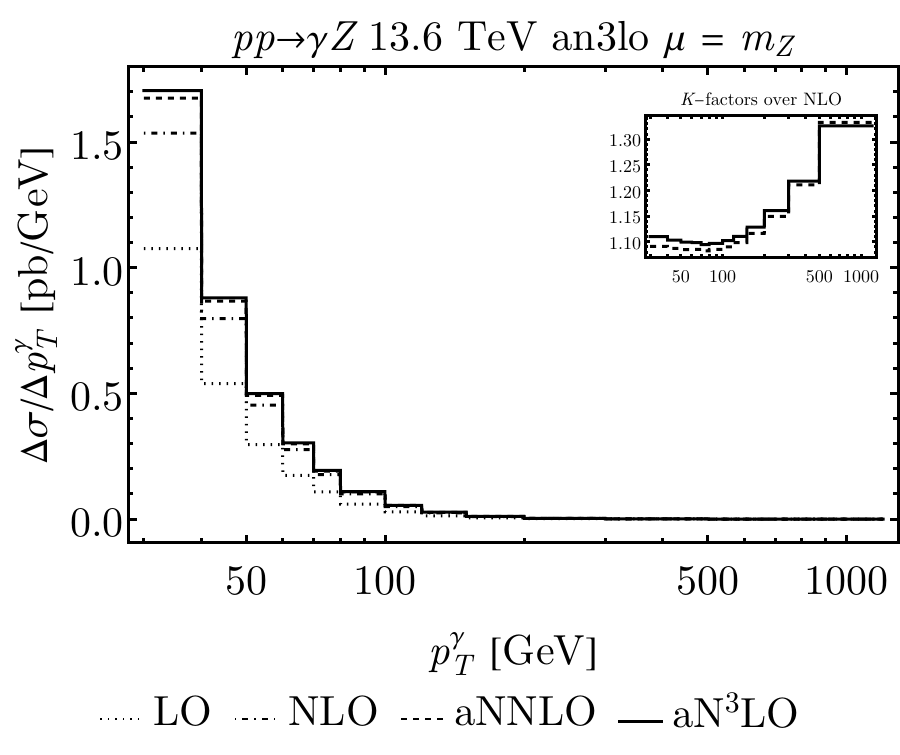}
    \includegraphics[width=0.4875\linewidth]{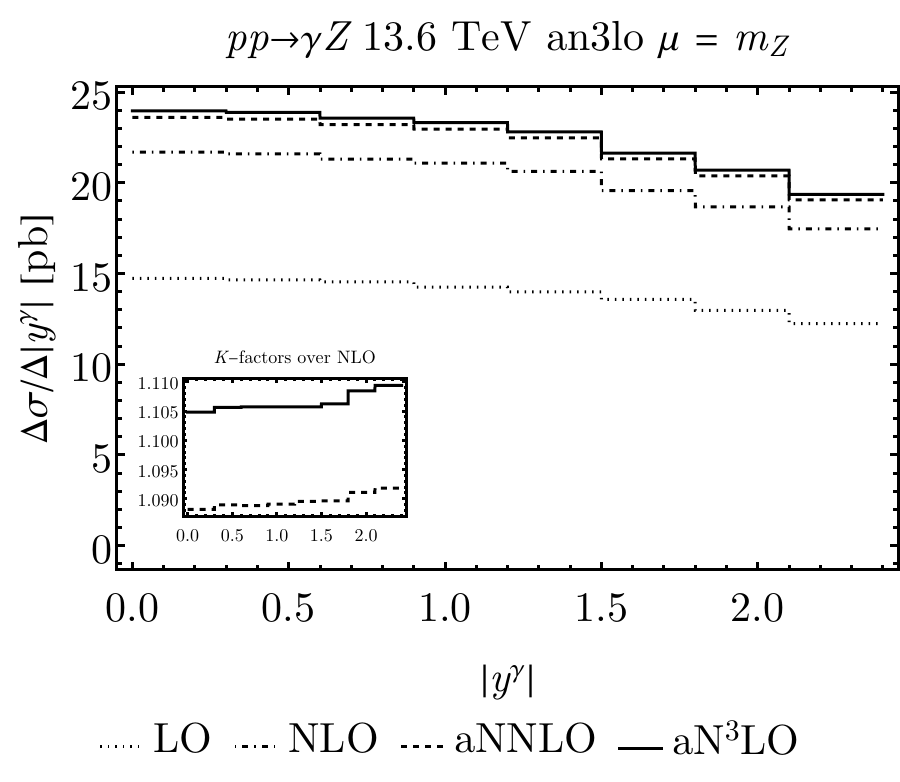}
    \caption{The same as Figure~\ref{fig:comparative_distributions_13_an3lo} but at 13.6~TeV.}
    \label{fig:comparative_distributions_13.6_an3lo}
\end{figure}

Figures~\ref{fig:comparative_distributions_13_nnlo}--\ref{fig:comparative_distributions_13.6_an3lo} show the $p_T^\gamma$ and $|y^\gamma|$ distributions at LO, NLO QCD, aNNLO QCD, and aN$^3$LO QCD at 13 and 13.6~TeV with nnlo and an3lo PDFs. Each panel includes an inset displaying the $K$-factors relative to NLO QCD for the aNNLO and aN$^3$LO predictions.


The $K$-factors relative to NLO QCD show distinct behavior across the two observables. For $p_T^\gamma$, the aNNLO/NLO and aN$^3$LO/NLO $K$-factors rise significantly at higher values of the $p_T^\gamma$ spectrum: they increase from roughly 1.09 and 1.11 in the lowest bin to approximately 1.33 at high $p_T^\gamma$, reflecting the growing importance of soft-gluon corrections at large transverse momenta where the threshold approximation becomes increasingly accurate. For $|y^\gamma|$, the $K$-factors are nearly uniform across the rapidity range: the aNNLO/NLO ratio is approximately 1.088--1.095 across all bins, while the aN$^3$LO/NLO ratio is approximately 1.104--1.112, with both ratios increasing mildly toward forward rapidities. These patterns are stable across both collider energies and PDF sets, with differences between 13 and 13.6~TeV and between nnlo and an3lo PDFs confined to at most 0.003 in any given bin. The contrast between the two observables reflects the dominance of soft-gluon corrections at large transverse momenta, while the rapidity distribution receives more uniform corrections across the full range.

\mysection{Conclusions}

In this paper, we computed soft-gluon corrections for $Z\gamma$ production through N$^3$LO in one-particle-inclusive kinematics. We combined these corrections with fixed-order LO and NLO predictions from {\small \sc MadGraph5\_aMC@NLO} to obtain aNNLO and aN$^3$LO differential cross sections for the photon transverse momentum and rapidity at collider energies of 13 TeV and 13.6 TeV, using both the MSHT20 nnlo and an3lo PDF sets.

Soft-gluon corrections are important across the full $p_T^\gamma$ range and grow larger at high $p_T^\gamma$. The aNNLO and aN$^3$LO $K$-factors relative to NLO rise from about 1.09 and 1.11 in the lowest bin to about 1.33 in the high-$p_T^\gamma$ tail. The $|y^\gamma|$ distribution receives more uniform corrections, with $K$-factors close to 1.09 at aNNLO and 1.11 at aN$^3$LO across the full rapidity range. The difference between the aNNLO and aN$^3$LO results is small in both observables. This shows good convergence of the perturbative series.

Scale uncertainties drop sharply from LO to NLO and then grow again at aNNLO and aN$^3$LO. This growth comes from the quadrature addition of the scale uncertainty of each new soft-gluon term. Even so, the aN$^3$LO scale uncertainty stays well below the LO uncertainty. PDF uncertainties stay close to 1.2 to 1.3 percent at every order. Our predictions change very little between the two PDF sets and between the two collider energies. This confirms that our results are stable.

These results give precise Standard Model predictions for $Z\gamma$ production at the LHC. They can serve as a baseline for measurements of anomalous neutral triple gauge couplings and for searches for physics beyond the Standard Model in the $Z\gamma$ final state. 

\section*{Acknowledgements}

This material is based upon work supported by the National Science Foundation under Grant No. PHY 2412071. K\c{S} was supported by the Kennesaw State University Office of Research Postdoctoral Fellowship Program. This work was supported in part by research computing resources and technical expertise via a partnership between Kennesaw State University's Office of the Vice President for Research and the Office of the CIO and Vice President for Information Technology. 

\appendix

\titleformat{\section}
  {\normalfont\large\bfseries}
  {Appendix}
  {1em}
  {}

\renewcommand{\theequation}{\Alph{section}.\arabic{equation}}
\counterwithin{equation}{section}

\section{N$^3$LO soft-gluon corrections for $\gamma Z$ production in 1PI kinematics}

Here we give explicit results for the soft-gluon contributions to the perturbative QCD corrections through N$^3$LO for $\gamma Z$ production in 1PI kinematics with kinematical variables as defined in Section II. We write the $n$th-order soft-plus-virtual (S+V) contributions to the differential partonic cross section as
\beq
\frac{d{\hat{\sigma}}^{(n) \, {\rm S+V}}_{q{\bar q} \to \gamma Z}}{dt_1 \, du_1} 
= F^{(0)}_{q{\bar q} \to \gamma Z} \left(\frac{\alpha_s(\mu_R)}{\pi}\right)^n \; 
\left\{ \sum_{k=0}^{2n-1} C^{(n)}_k \left[\frac{\ln^k(s_4/m_Z^2)}{s_4}\right]_+ 
+C^{(n)}_{\delta} \, \delta(s_4) \right\}
\label{nsoft}
\eeq
where the coefficients $C^{(n)}_k$ are given below through N$^3$LO. 

The NLO coefficients are given by
\beq
C^{(1)}_1=4 C_F \, ,
\eeq
\beq
C^{(1)}_0=-2 \, C_F \, \ln\left(\frac{t_1 u_1}{m_Z^4}\right) -2 \, C_F \, \ln\left(\frac{\mu_F^2}{s}\right) \, ,
\eeq
and 
\beq
C^{(1)}_{\delta}=-4 C_F + 2C_F \zeta_2+C_F\, \ln^2\left(\frac{-t_1}{m_Z^2}\right)
+ C_F\, \ln^2\left(\frac{-u_1}{m_Z^2}\right)
+\left[C_F \, \ln\left(\frac{t_1 u_1}{m_Z^4}\right)
-\frac{3}{2}C_F\right]\ln\left(\frac{\mu_F^2}{s}\right) \, .
\eeq
Here $C_F=(N_c^2-1)/(2 N_c)$ with $N_c=3$ the number of colors, and $\zeta_2=\pi^2/6$.

The NNLO coefficients are given by
\beq
C^{(2)}_3=8 C_F^2 \, ,
\eeq
\beq
C^{(2)}_2=-\frac{11}{3} C_F C_A+\frac{2}{3} C_F n_f
-12 C_F^2 \ln\left(\frac{t_1 u_1}{m_Z^4}\right)
-12 C_F^2 \ln\left(\frac{\mu_F^2}{s}\right) \, ,
\eeq
\beqa
C^{(2)}_1 &=& -16 C_F^2-8 C_F^2\zeta_2+\frac{67}{9} C_F C_A-2 C_F C_A \zeta_2-\frac{10}{9} C_F n_f
+4 C_F^2 \ln^2\left(\frac{-t_1}{m_Z^2}\right)+4 C_F^2 \ln^2\left(\frac{-u_1}{m_Z^2}\right)
\nonumber \\ && 
{}+4 C_F^2 \ln^2\left(\frac{t_1 u_1}{m_Z^4}\right)+\left(\frac{11}{3} C_F C_A -\frac{2}{3} C_F n_f \right) \ln\left(\frac{t_1 u_1}{m_Z^4}\right)
+4 C_F^2 \ln^2\left(\frac{\mu_F^2}{s}\right)
\nonumber \\ && 
{}-6 C_F^2 \ln\left(\frac{\mu_F^2}{s}\right)
+12 C_F^2 \ln\left(\frac{t_1 u_1}{m_Z^4}\right) \ln\left(\frac{\mu_F^2}{s}\right) 
+\left(\frac{11}{3} C_F C_A -\frac{2}{3} C_F n_f \right) \ln\left(\frac{\mu_R^2}{s}\right) \, ,
\eeqa
and
\beqa
C^{(2)}_0 &=& 16 C_F^2 \zeta_3+C_F C_A \left(-\frac{101}{27} + \frac{11}{3} \zeta_2 + \frac{7}{2} \zeta_3\right) + C_F n_f \left(\frac{14}{27} - \frac{2}{3} \zeta_2\right)
\nonumber \\ &&
{}-2 C_F^2 \ln^2\left(\frac{-t_1}{m_Z^2}\right) \ln\left(\frac{t_1 u_1}{m_Z^4}\right)
-2 C_F^2 \ln^2\left(\frac{-u_1}{m_Z^2}\right) \ln\left(\frac{t_1 u_1}{m_Z^4}\right)
\nonumber \\ && 
{}-\left(\frac{11}{6} C_F C_A-C_F \frac{n_f}{3}\right) \ln^2\left(\frac{-t_1}{m_Z^2}\right)
-\left(\frac{11}{6} C_F C_A-C_F \frac{n_f}{3}\right) \ln^2\left(\frac{-u_1}{m_Z^2}\right)
\nonumber \\ && 
{}+\left(8C_F^2+4 C_F^2 \zeta_2 -\frac{67}{18} C_F C_A+C_F C_A \zeta_2+\frac{5}{9} C_F n_f \right) 
\ln\left(\frac{t_1 u_1}{m_Z^4}\right)
\nonumber \\ && 
{}+\left(3C_F^2 +\frac{11}{12} C_F C_A- C_F \frac{n_f}{6}\right) \ln^2\left(\frac{\mu_F^2}{s}\right)
-2 C_F^2 \ln\left(\frac{t_1 u_1}{m_Z^4}\right) \ln^2\left(\frac{\mu_F^2}{s}\right)
\nonumber \\ &&
{}-\left(\frac{11}{6} C_F C_A-C_F \frac{n_f}{3}\right) \ln\left(\frac{\mu_F^2}{s}\right) \ln\left(\frac{\mu_R^2}{s}\right)
\nonumber \\ && 
{}+\left(8C_F^2+4 C_F^2 \zeta_2 -\frac{67}{18} C_F C_A+C_F C_A \zeta_2+\frac{5}{9} C_F n_f \right) 
\ln\left(\frac{\mu_F^2}{s}\right)
\nonumber \\ &&
{}-2 C_F^2 \ln^2\left(\frac{-t_1}{m_Z^2}\right) \ln\left(\frac{\mu_F^2}{s}\right)
-2 C_F^2 \ln^2\left(\frac{-u_1}{m_Z^2}\right) \ln\left(\frac{\mu_F^2}{s}\right)
-2 C_F^2 \ln^2\left(\frac{t_1 u_1}{m_Z^4}\right) \ln\left(\frac{\mu_F^2}{s}\right)
\nonumber \\ && 
{}+3C_F^2 \ln\left(\frac{t_1 u_1}{m_Z^4}\right) \ln\left(\frac{\mu_F^2}{s}\right)
-\left(\frac{11}{6} C_F C_A-C_F \frac{n_f}{3}\right) \ln\left(\frac{t_1 u_1}{m_Z^4}\right) \ln\left(\frac{\mu_R^2}{s}\right)
\, ,
\label{NNLOC210}
\eeqa
while the expression for $C^{(2)}_{\delta}$ is quite long and is given as supplementary material. Here $C_A=N_c$, $n_f$ is the number of light-quark flavors (set equal to 5 in our calculations), and $\zeta_3=1.202056903\cdots$.

The N$^3$LO coefficients are given by
\beq
C^{(3)}_5=8 C_F^3 \, ,
\eeq
\beq
C^{(3)}_4=-\frac{110}{9} C_F^2 C_A +\frac{20}{9} C_F^2 n_f -20 C_F^3 \ln\left(\frac{t_1 u_1}{m_Z^4}\right)
-20 C_F^3 \ln\left(\frac{\mu_F^2}{s}\right) \, ,
\eeq
\beqa
C^{(3)}_3&=& -32 C_F^3-48 C_F^3 \zeta_2+\frac{268}{9} C_F^2 C_A-8 C_F^2 C_A \zeta_2+\frac{121}{27} C_F C_A^2-\frac{40}{9} C_F^2 n_f-\frac{44}{27} C_F C_A n_f
\nonumber \\ &&
{}+\frac{4}{27} C_F n_f^2
+8 C_F^3 \ln^2\left(\frac{-t_1}{m_Z^2}\right)+8 C_F^3 \ln^2\left(\frac{-u_1}{m_Z^2}\right)
+16 C_F^3 \ln^2\left(\frac{t_1 u_1}{m_Z^4}\right)
\nonumber \\ &&
{}+\left(\frac{220}{9} C_F^2 C_A -\frac{40}{9} C_F^2 n_f \right) \ln\left(\frac{t_1 u_1}{m_Z^4}\right)
+16 C_F^3  \ln^2\left(\frac{\mu_F^2}{s}\right)
\nonumber \\ &&
{}+\left(-12C_F^3+\frac{88}{9} C_F^2 C_A -\frac{16}{9} C_F^2 n_f\right) \ln\left(\frac{\mu_F^2}{s}\right)
+40 C_F^3 \ln\left(\frac{t_1 u_1}{m_Z^4}\right) \ln\left(\frac{\mu_F^2}{s}\right)
\nonumber \\ &&
{}+\left(\frac{44}{3} C_F^2 C_A -\frac{8}{3} C_F^2 n_f \right) \ln\left(\frac{\mu_R^2}{s}\right) \, ,
\eeqa
\beqa
C^{(3)}_2&=& 160 C_F^3 \zeta_3 - \frac{70}{9} C_F^2 C_A + \frac{176}{3} C_F^2 C_A \zeta_2 
+ 21 C_F^2 C_A \zeta_3 -\frac{445}{27} C_F C_A^2 + \frac{11}{3} C_F C_A^2 \zeta_2 
\nonumber \\ &&
{}+ \frac{17}{18} C_F^2 n_f - \frac{32}{3} C_F^2 n_f \zeta_2 + \frac{289}{54} C_F C_A n_f 
- \frac{2}{3} C_F C_A n_f \zeta_2 - \frac{10}{27} C_F n_f^2 
\nonumber \\ &&
{}-12 C_F^3 \ln^2\left(\frac{-t_1}{m_Z^2}\right) \ln\left(\frac{t_1 u_1}{m_Z^4}\right)
-12 C_F^3 \ln^2\left(\frac{-u_1}{m_Z^2}\right) \ln\left(\frac{t_1 u_1}{m_Z^4}\right)
-4 C_F^3 \ln^3\left(\frac{t_1 u_1}{m_Z^4}\right) 
\nonumber \\ &&
{}-\left(\frac{44}{3} C_F^2 C_A -\frac{8}{3} C_F^2 n_f \right) \ln^2\left(\frac{-t_1}{m_Z^2}\right)
-\left(\frac{44}{3} C_F^2 C_A -\frac{8}{3} C_F^2 n_f \right) \ln^2\left(\frac{-u_1}{m_Z^2}\right)
\nonumber \\ &&
{}-\left(11 C_F^2 C_A - 2 C_F^2 n_f \right) \ln^2\left(\frac{t_1 u_1}{m_Z^4}\right)
+\left(48 C_F^3+72 C_F^3 \zeta_2-\frac{134}{3} C_F^2 C_A + 12 C_F^2 C_A \zeta_2 \right.
\nonumber \\ && \hspace{10mm} \left.
{} -\frac{121}{18} C_F C_A^2+\frac{20}{3} C_F^2 n_f+\frac{22}{9} C_F C_A n_f-\frac{2}{9} C_F n_f^2 \right) \ln\left(\frac{t_1 u_1}{m_Z^4}\right)
\nonumber \\ &&
{}-4 C_F^3 \ln^3\left(\frac{\mu_F^2}{s}\right)+\left(18C_F^3+\frac{11}{2} C_F^2 C_A - C_F^2 n_f \right) \ln^2\left(\frac{\mu_F^2}{s}\right)
\nonumber \\ &&
{}-24 C_F^3 \ln\left(\frac{t_1 u_1}{m_Z^4}\right)  \ln^2\left(\frac{\mu_F^2}{s}\right) 
-\left(22 C_F^2 C_A -4C_F^2 n_f \right) \ln\left(\frac{\mu_F^2}{s}\right) \ln\left(\frac{\mu_R^2}{s}\right)
\nonumber \\ &&
{}+\left(48 C_F^3+72 C_F^3 \zeta_2-\frac{235}{6} C_F^2 C_A +12 C_F^2 C_A \zeta_2+\frac{17}{3}C_F^2 n_f \right) \ln\left(\frac{\mu_F^2}{s}\right)
\nonumber \\ &&
{}-12 C_F^3 \ln^2\left(\frac{-t_1}{m_Z^2}\right) \ln\left(\frac{\mu_F^2}{s}\right) 
-12 C_F^3 \ln^2\left(\frac{-u_1}{m_Z^2}\right) \ln\left(\frac{\mu_F^2}{s}\right)
-24 C_F^3 \ln^2\left(\frac{t_1 u_1}{m_Z^4}\right) \ln\left(\frac{\mu_F^2}{s}\right)
\nonumber \\ &&
{}+\left(18 C_F^3-\frac{44}{3} C_F^2 C_A +\frac{8}{3} C_F^2 n_f \right) \ln\left(\frac{t_1 u_1}{m_Z^4}\right) \ln\left(\frac{\mu_F^2}{s}\right)
\nonumber \\ &&
{}-\left(\frac{121}{18} C_F C_A^2 -\frac{22}{9} C_F C_A n_f+\frac{2}{9} C_F n_f^2\right) \ln\left(\frac{\mu_R^2}{s}\right)
\nonumber \\ &&
{}-\left(22 C_F^2 C_A -4 C_F^2 n_f \right) \ln\left(\frac{t_1 u_1}{m_Z^4}\right) \ln\left(\frac{\mu_R^2}{s}\right) \, ,
\eeqa
and $C^{(3)}_1$ and $C^{(3)}_0$ have very long expressions which are given as supplementary material.


\begin{thebibliography}{99}

\bibitem{ATLAS2018}
ATLAS Collaboration, Measurement of the $Z \gamma \to \nu {\bar \nu} \gamma$ production cross section in $pp$ collisions at ${\sqrt s} = 13$ TeV with the ATLAS detector and limits on anomalous triple gauge-boson couplings, JHEP {\bf 12}, 010 (2018) [arXiv:1810.04995].

\bibitem{CMS2026}
CMS Collaboration, Measurement of the $Z \gamma$ production cross section and search for anomalous neutral triple gauge couplings in $pp$ collisions at ${\sqrt s} = 13$ TeV, arXiv:2601.14102.

\bibitem{Degrande2013}
C. Degrande, A basis of dimension-eight operators for anomalous neutral triple gauge boson interactions, JHEP {\bf 02}, 101 (2014) [arXiv:1308.6323].

\bibitem{Ohnemus1993}
J. Ohnemus, Order-$\alpha_s$ calculations of hadronic $W^{\pm} \gamma$ and $Z \gamma$ production, Phys. Rev. D {\bf 47}, 940 (1993).

\bibitem{GKRT2013}
M. Grazzini, S. Kallweit, D. Rathlev, and A. Torre, $Z \gamma$ production at hadron colliders in NNLO QCD, Phys. Lett. B {\bf 731}, 204 (2014) [arXiv:1309.7000].

\bibitem{GKR2015}
M. Grazzini, S. Kallweit, and D. Rathlev, $W \gamma$ and $Z \gamma$ production at the LHC in NNLO QCD, JHEP {\bf 07}, 085 (2015) [arXiv:1504.01330].

\bibitem{ATLAS2019}
ATLAS Collaboration, Measurement of the $Z(\to l^+ l^-) \gamma$ production cross-section in $pp$ collisions at $\sqrt s = 13$ TeV with the ATLAS detector, JHEP {\bf 03}, 054 (2020) [arXiv:1911.04813].

\bibitem{Sterman1987}
G. Sterman, Summation of large corrections to short-distance hadronic cross sections, Nucl. Phys. B {\bf 281}, 310 (1987).

\bibitem{NKGS1996}
N. Kidonakis and G. Sterman, Subleading logarithms in QCD hard scattering, Phys. Lett. B {\bf 387}, 867 (1996).

\bibitem{NKGS1997}
N. Kidonakis and G. Sterman, Resummation for QCD hard scattering, Nucl. Phys. B {\bf 505}, 321 (1997) [arXiv:hep-ph/9705234].

\bibitem{KOS1998}
N. Kidonakis, G. Oderda, and G. Sterman, Evolution of color exchange in QCD hard scattering, Nucl. Phys. B {\bf 531}, 365 (1998) [arXiv:hep-ph/9803241].

\bibitem{LOS1998}
E. Laenen, G. Oderda, and G. Sterman, Resummation of threshold corrections for single-particle inclusive cross sections, Phys. Lett. B {\bf 438}, 173 (1998) [arXiv:hep-ph/9806467].

\bibitem{Kidonakis2007}
N. Kidonakis, Collinear and soft gluon corrections to Higgs production at next-to-next-to-next-to-leading order, Phys. Rev. D {\bf 77}, 053008 (2008) [arXiv:0711.0142].

\bibitem{Kidonakis2010}
N. Kidonakis, Two-loop soft anomalous dimensions for single top quark associated production with a $W^-$ or $H^-$ , Phys. Rev. D {\bf 82}, 054018 (2010) [arXiv:1005.4451].

\bibitem{NKRG2014}
N. Kidonakis and R.J. Gonsalves, NNLO soft-gluon corrections for the $Z$ boson and $W$ boson transverse momentum distributions, Phys. Rev. D {\bf 89}, 094022 (2014) [arXiv:1404.4302]. 

\bibitem{Kidonakis2017}
N. Kidonakis, Higher-order corrections for $tZ$ production via anomalous couplings, Phys. Rev. D {\bf 97}, 034028 (2018) [arXiv:1712.01144].

\bibitem{MFNK2020}
M. Forslund and N. Kidonakis, Resummation for $2 \to n$ processes in single-particle-inclusive kinematics, Phys. Rev. D {\bf 102}, 034006 (2020) [arXiv:2003.09021].

\bibitem{NKNY2022}
N. Kidonakis and N. Yamanaka, Soft-gluon corrections for $tqZ$ production, Phys. Lett. B {\bf 838}, 137708 (2023) [arXiv:2210.09542].

\bibitem{NKAT2024}
N. Kidonakis and A. Tonero, N$^3$LO soft-gluon corrections in single-particle-inclusive kinematics and $H^+ H^-$ production, JHEP {\bf 06}, 138 (2024) [arXiv:2404.00089].

\bibitem{NKCF2024}
N. Kidonakis and C. Foster, Higher-order soft-gluon corrections for $t{\bar t}Z$ cross sections, Phys. Lett. B {\bf 860}, 139146 (2025) [arXiv:2410.01214].

\bibitem{NKAT2025}
N. Kidonakis and A. Tonero, Higher-order soft and virtual corrections in $pp \to \gamma W$ production at the LHC, Eur. Phys. J. C {\bf 85}, 1270 (2025) [arXiv:2506.01590].

\bibitem{Alwalletal2014}
J. Alwall {\sl et al.}, The automated computation of tree-level and next-to-leading order differential cross sections, and their matching to parton shower simulations, JHEP {\bf 07}, 079 (2014) [arXiv:1405.0301].

\bibitem{Frederixetal2018}
R. Frederix {\sl et al.}, The automation of next-to-leading order electroweak calculations, JHEP {\bf 07}, 185 (2018) [Erratum: JHEP {\bf 11}, 085 (2021)] [arXiv:1804.10017].

\bibitem{Frixione1998}
S. Frixione, Isolated photons in perturbative QCD, Phys. Lett. B {\bf 429}, 369 (1998) [arXiv:hep-ph/9801442].

\bibitem{MSHT20nnlo}
S. Bailey, T. Cridge, L.A. Harland-Lang, A.D. Martin, and R.S. Thorne, Parton distributions from LHC, HERA, Tevatron and fixed target data: MSHT20 PDFs, Eur. Phys. J. C {\bf 81}, 341 (2021) [arXiv:2012.04684].

\bibitem{MSHT20an3lo}
J. McGowan, T. Cridge, L.A. Harland-Lang, and R.S. Thorne, Approximate N$^3$LO parton distribution functions with theoretical uncertainties: MSHT20aN$^3$LO PDFs, Eur. Phys. J. C {\bf 83}, 185 (2023) [arXiv:2207.04739].

\end{thebibliography}
\end{document}